\documentclass[reprint,aps,pra,superscriptaddress]{revtex4-2}

\usepackage{float}
\usepackage{soul}
\usepackage{graphicx}
\usepackage{dcolumn}
\usepackage{bm}
\usepackage[colorlinks=true,allcolors=blue]{hyperref}
\usepackage[utf8]{inputenc}
\usepackage[english]{babel}
\usepackage{hyperref} 
\usepackage{booktabs}
\usepackage{colortbl}
\usepackage{multirow}
\usepackage{amsmath, amsthm, amscd, amssymb} 
\usepackage[mathlines]{lineno}

\begin{document}

\title{Vacuum levitation and motion control on chip}

\author{Bruno Melo}
\thanks{equal contribution}
\affiliation{Nanophotonic Systems Laboratory, Department of Mechanical and Process Engineering, ETH Zurich, 8092 Zurich, Switzerland}
\affiliation{Quantum Center, ETH Zurich, 8083 Zurich, Switzerland}
\author{Marc T. Cuairan}
\thanks{equal contribution}
\affiliation{Nanophotonic Systems Laboratory, Department of Mechanical and Process Engineering, ETH Zurich, 8092 Zurich, Switzerland}
\affiliation{Quantum Center, ETH Zurich, 8083 Zurich, Switzerland}

\author{Grégoire F. M. Tomassi}
\thanks{}
\affiliation{Nanophotonic Systems Laboratory, Department of Mechanical and Process Engineering, ETH Zurich, 8092 Zurich, Switzerland}
\affiliation{Quantum Center, ETH Zurich, 8083 Zurich, Switzerland}
\author{Nadine Meyer}
\email{{nmeyer@ethz.ch}}
\affiliation{Nanophotonic Systems Laboratory, Department of Mechanical and Process Engineering, ETH Zurich, 8092 Zurich, Switzerland}
\affiliation{Quantum Center, ETH Zurich, 8083 Zurich, Switzerland}
\email{nmeyer@ethz.ch}
\author{Romain Quidant}
\email{rquidant@ethz.ch}
\affiliation{Nanophotonic Systems Laboratory, Department of Mechanical and Process Engineering, ETH Zurich, 8092 Zurich, Switzerland}
\affiliation{Quantum Center, ETH Zurich, 8083 Zurich, Switzerland}

\date{\today}

\maketitle

\textbf{Levitation in vacuum has evolved into a versatile technique which has already benefited diverse scientific directions, from force sensing and thermodynamics to material science and chemistry. It also holds great promises of advancing the study of quantum mechanics in the unexplored macroscopic regime. While most current levitation platforms are complex and bulky, miniaturization is sought to gain robustness and facilitate their integration into confined settings, such as cryostats or portable devices. Integration on chip is also anticipated to enhance the control over the particle motion through a more precise engineering of optical and electric fields. As a substantial milestone towards this goal, we present here levitation and motion control in high vacuum of a silica nanoparticle at the surface of a hybrid optical-electrostatic chip. By combining fiber-based optical trapping and sensitive position detection with cold damping through planar electrodes, we cool the particle motion to a few hundred phonons. Our results pave the way to the next generation of integrated levitation platforms combining integrated photonics and nanophotonics with engineered electric potentials, towards complex state preparation and read out.}



Since the seminal experiment by Ashkin \textit{et al.}, reporting on the levitation in vacuum of a microsphere~\cite{ashkin1971optical}, much progress has been made in controlling the translational and rotational degrees of freedom of levitated objects~\cite{gonzalez2021levitodynamics}. While most efforts have focused on purely optical approaches, more recent developments evolved towards hybrid platforms combining techniques adapted from atomic physics. For instance, to overcome the constraints posed by intense optical fields in terms of both photo-damage and recoil heating~\cite{jain_direct_2016}, researchers have introduced hybrid optical/electric potentials which combine spatial confinement with high potential depth~\cite{millen2015cavity,conangla2020extending,bykov2022hybrid}. Furthermore, linear feedback using electric forces on a charged particle~\cite{Conangla_OptimalFB,Tebbenjohanns_ColdDamp} enables more efficient cooling compared to its parametric counterpart~\cite{gieseler_subkelvin_2012}. Leveraging the palette of available techniques developed over the last decade has enabled the achievement of important milestones, including precision acceleration~\cite{timberlake2019acceleration, Monteiro_force_acceleration_sensing_2020} force~\cite{winstone_electrostaticimage_force_sensing,Hempston_force,ranjit2016zeptonewton,Kawasaki_short-distance_2020,Hebestreit_staticforce} and torque~\cite{ahn2020ultrasensitive} sensing, as well as reaching ground-state cooling in both  one-~\cite{magrini_real-time_2021,tebbenjohanns_quantum_2021, delic2020cooling,ranfagni2022two,kamba2022optical}  and two-dimensions \cite{piotrowski2023simultaneous}.\\

As it gains maturity, vacuum levitation is now entering a phase of miniaturization, following the footsteps of trapped ions~\cite{bruzewicz2019trapped} and solid-state ~\cite{kjaergaard2020superconducting} systems. Beyond reducing bulkiness and mitigating instabilities associated to multiple assembled parts, integration on chip is envisioned to facilitate advancements essential for the next generations of experiments. Nanoscale engineering at the chip surface can first provide a finer control over the electric~\cite{madsen2004planar,pearson2006experimental,kim2010surface} and magnetic~\cite{latorre2022chip} fields experienced by the particle. It also offers the possibility to enhance the nanoparticle's interaction with optical fields by exploiting subwavelength modes supported by near field nanocavities~\cite{magrini2018near}. Interfacing with integrated photonics and meta-optics also has the potential to facilitate scalability towards arrays of multiple traps ~\cite{rieser2022tunable,vijayan2023scalable,vcivzmar2006optical}. Finally, on chip integration provides a pathway towards implementing on a single platform complex dynamical protocols involving bright and dark potentials~\cite{roda2023macroscopic}. While miniaturization is well underway, with examples such as levitation in planar ion traps~\cite{jin2023quantum,alda2016trapping,kim2010surface} and optical trapping at the focus of a metalens~\cite{shen2021chip}, fully integrated platforms are missing.


Here, we present a hybrid photonic-electric on-chip platform enabling the robust levitation, precise position detection and dynamical control of a nanoparticle in vacuum. Our approach circumvents the need for bulky high numerical aperture (NA) lenses by combining commercial optical fibers with micro-scale additive manufacturing to create a robust, versatile and flexible optical interface. Despite the absence of focusing optics, we reach high signal to noise ratios (SNR) in optical displacement detection that compete with bulky, high NA optics. When combined with active feedback cooling with planar electrodes, we efficiently cool down the particle motion in three dimensions. \\ 

\begin{figure}
\centering
\includegraphics[width=0.45\textwidth]{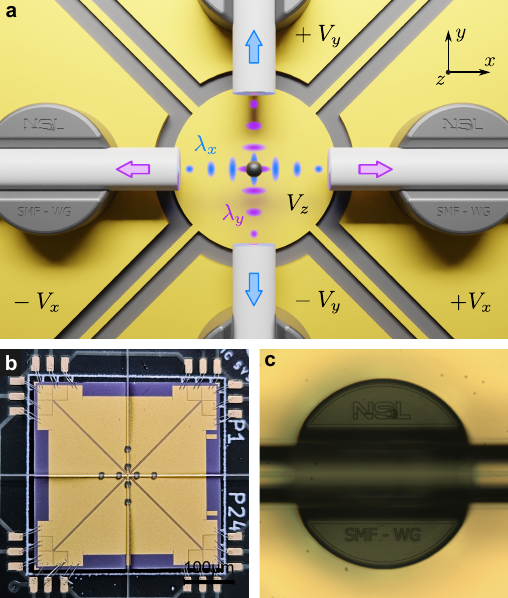}
\caption{\textbf{On chip levitation platform:} \textbf{a)} The upper optical layer consists of two orthogonal pairs of cleaved single mode optical fibers. One of the pairs  (along $y$) creates a standing wave at $\lambda_y=1550~\text{nm}$, while the second pair (along $x$) creates a standing wave at $\lambda_x=1064~\text{nm}$. A particle (black) is trapped at the intersection of both standing waves. The light scattered by the particle into the fibers, represented by the arrows, is used for displacement detection. The four fibers are positioned above a set of planar electrodes used to apply active feedback cooling to the charged particle via electric forces: right and left electrodes for feedback along $x$, top and bottom for feedback along $y$ and center electrode for feedback along $z$. \textbf{b)} Picture of the levitation chip showing the planar electrodes, four optical fibers, fiber mounts close to the center and wire bonds from the chip to the PCB at the corners. \textbf{c)} Optical fiber positioned into a mechanical mount fabricated via two-photon polymerization and used to align and hold the fibers in place.
}
\label{fig:1}
\end{figure}

Our levitation chip is structured in two layers: an upper photonic layer, where the particle is trapped and that allows precise detection of the nanoparticle's motion through the analysis of scattered light, and a lower electric layer formed by a set of planar electrodes to cool the particle's motion. Facilitating direct fiber interfacing, the photonic layer consists of an arrangement of four orthogonal cleaved single mode optical fibers as illustrated in Fig.~\ref{fig:1}a. Trapping of the particle occurs in the standing wave pattern formed by two interfering counter-propagating beams~\cite{ zemanek1998optical, kamba2021recoil}. This configuration has the combined advantage to efficiently cancel the scattering force while creating multiple trapping sites. 

Let us first consider one single standing wave (SW) along $y$ formed by two equally linearly polarized counter-propagating divergent beams as emerging from two single mode fibers of numerical aperture NA = 0.1~\cite{ashkin1970acceleration,constable1993demonstration} , separated by a distance of $d=160~\mu\text{m}$. The wavelength is $\lambda_y=1550~\text{nm}$ and the total power $P=1$ W. The equally linearly polarised light field interacts with a nanoparticle of refractive index $n_r$, radius $R=160~\text{nm}$ and polarizability $\alpha = 4\pi \epsilon_0 R^3 (n_r^2-1)/(n_r^2+2)$. At each intensity antinode, the optical force experienced by the particle gives rise to a harmonic potential with theoretical mechanical eigen-frequencies $\Omega_{x,y,z}/(2\pi) \approx(3.5,89,3.5)$ kHz and a trap depth $U= 42 k_BT_0$ where $k_B$ is the Boltzmann constant and $T_0= 300$K room temperature. 

In order to achieve 3D active feedback stabilization, it is necessary to ensure well-separated mechanical frequencies along each axis ~\cite{vijayan2023scalable}. To this aim, we add a second SW with a wavelength $\lambda_x= 1064$ nm along the $x$-axis, as shown in Fig.~\ref{fig:1}a. The combination of both SWs results in a high-frequency mechanical mode along each optical axis ($x$ and $y$) and a low-frequency mechanical mode ($z$) along the vertical axis.

Beyond ensuring robust trapping above the chip surface and an accurate adjustment of mechanical frequencies, the two pairs of optical fibers serve the additional purpose of  monitoring the particle's position by detecting its scattering. Exploiting the access to both optical axis, the scattered light from each SW is collected by the orthogonal fiber pair (see arrows in Fig.~\ref{fig:1}a) and used to monitor the center of mass (COM) motion.
This distinctive collection scheme is inaccessible in single beam traps and has the advantage to better adjust to the scattering pattern of the particle~\cite{maurer2023quantum}.\\

\begin{figure}
\centering
\includegraphics[width=0.45\textwidth]{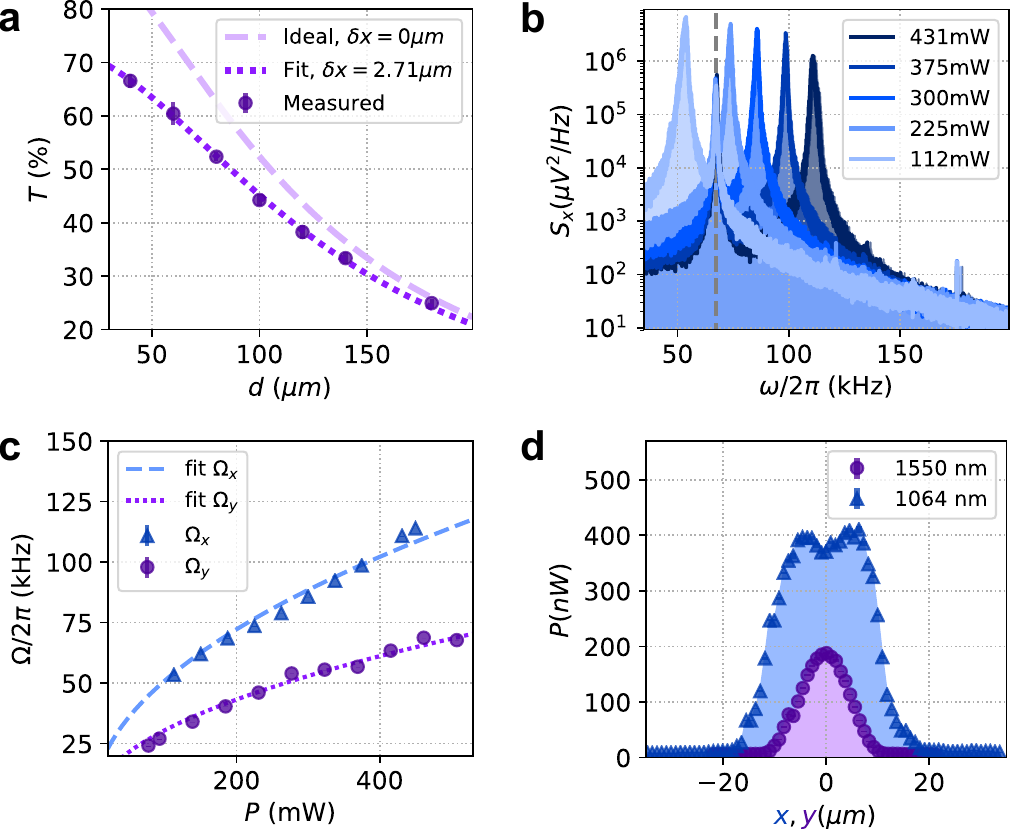}
\caption{\textbf{Characterization of the fiber-based trap} \textbf{a)} Transmission $T$ between two fibers held by the fiber mounts as a function of the distance $d$ between the fibers end facets. A misalignment of $\delta x\approx2.71\pm0.03\mu\text{m}$ between the fibers is estimated by fitting the measured transmission (circles) to a theoretical model (dotted line, see SM-II). The dashed line shows the theoretical model in absence of misalignment. \textbf{b)} PSDs of the motion along $x$ for different powers $P_x$ of the 1064 nm light, while maintaining the 1550~\text{nm} laser power $P_y$ constant. The trapping powers are estimated from the input power and the coupling and transmission efficiencies throughout the setup. \textbf{c)} Mechanical eigenfrequencies $\Omega_{x,y}$ of the motion along $x$ (blue triangles) and $y$ (purple circles) for different powers of the respective trapping beam. The lines show the fit to the expected behaviour $\Omega_j\propto\sqrt{P_j}$. \textbf{d)} Power collected by the fibers for different positions ($x,y$) of the particle relative to the center of the collecting fiber. The 1550nm (1064nm) scattered light is collected by a fiber along $x$ ($y$) while the particle is moved along $y$ ($x$). The $0$ position is defined as the approximate center of the curves.}
\label{fig:2}
\end{figure}

The fabricated chip measures $0.5~\text{in}\times 0.5~\text{in}$ and is mounted on a custom-made printed circuit board (PCB) for electrical interfacing, see Fig.~\ref{fig:1}b. The electrostatic layer consists of five planar electrodes (see Fig.~\ref{fig:1}a and c), which are used to apply active electrical feedback via electric fields. 
To achieve reliable levitation, precise control over the position of each cleaved optical fiber is of upmost importance. Consequently, each individual fiber is hold in two U-shaped mechanical mounts as displayed in Fig.~\ref{fig:1}c. The latter are microfabricated via two-photon polymerization with a commercial Nanoscribe device ~\cite{yu2023free, wang2023two} (see SM-I for further details). The relative fiber alignment is assessed by measuring the transmission $T$ from fiber to fiber (Thorlabs SMF-28), as shown for 1550nm light for different fiber separations $d$ in Fig.~\ref{fig:2}a. By fitting the data (circles) to a theoretical model (see SM-II) we extract a relative fiber misalignment of $\delta x \approx 2.71\mu \text{m}$ (dotted line). The dashed line shows the ideal case of no misalignment. Once aligned, the cleaved fibers are fixed with epoxy. During the curing process, the transmission $T$ varies by a few percent, without significant long term drifts at constant pressure (see SM-II). Nevertheless, $T$ consistently increases by around 4$\%$ from ambient pressure to vacuum. In general, $T$ is stable in time, making this a reliable and robust method to position fibers permanently. Note that our fiber mounting method can be easily employed to create more complex optical lattices due to the arbitrary in-plane positioning of the fibers. 


To maximize the trap depth, it is beneficial to position the fiber facets as close as possible to each other. Along $y$, this is limited by the diameter of the fiber cladding  ($125 \mu \text{m}$, Thorlabs 1060XP). Along $x$, $d_x$ is limited by the diameter of the diverging 1550$\text{nm}$ beam. To have a safety margin on these constraints and to diminish the interference coming from reflections, we work with $d_y=160\mu \text{m}$ and $d_x=80\mu \text{m}$. No additional fiber treatment or optics are employed for optical trapping.  \\

Following standard nebulization-based loading, a single silica particle of radius $R\approx 160 \text{nm}$ is trapped at the intersection of the two SWs of wavelengths $\lambda_x = 1064\text{nm}$ and  $\lambda_y = 1550\text{nm}$. One distinctive feature of the intersecting SWs lies in the ability to independently tune the mechanical eigenfrequencies $\Omega_q$ with $q=x,y,z$ . As displayed in Fig.~\ref{fig:2}b, by decreasing the power $P_{x}$ of the 1064~\text{nm} SW, we observe the expected decrease of $\Omega_x$, while $\Omega_y$ remains constant (dashed line). Additionally, by independently varying $P_x$ and $P_y$ of each SW and extracting $\Omega_{x}$ (purple circles) and $\Omega_{y}$ (blue triangles), we verify in Fig.~\ref{fig:2}c the expected behavior as $\Omega_q\propto \sqrt{P_{q}}$ (dashed and dotted lines). Remarkably, despite the use of low NA fibers, the achieved values of $\Omega_q$ are comparable to high NA optics. 

The particle's position along $x,y$ is controlled by changing the relative phase $\phi_q$ between the corresponding counter-propagating beams (see SM-III). In Fig~\ref{fig:2}d, we move the particle along one axis and measure how much light is scattered into the fiber along the perpendicular direction. The blue triangles (purple circles) show the $1064\text{nm}$ ($1550\text{nm}$) power $P_x$ ($P_y$) collected by the fibers along the $y$ ($x$) axis. We attribute the shape of the $1064\text{nm}$ curve to the multimode character of the Thorlabs SMF-28 fiber at $\lambda_x$. To ensure high photon collection efficiency in both directions while maintaining symmetry, we place the particle in the position corresponding to $x,y=0$.

To achieve high feedback efficiency, it is important to consider the angular distribution of the motional information radiation pattern~\cite{tebbenjohanns2019optimal,maurer2023quantum}. For particles trapped by a single beam, most of the information about the particle's axial motion is contained in the back scattered light, enabling 1D groundstate cooling via measurement based feedback ~\cite{magrini_real-time_2021,tebbenjohanns_quantum_2021,kamba2022optical}. In contrast to the single beam configuration, here the second trapping beam of the SW acts as a strong local oscillator with fixed phase relation with respect to the back-scattered light. Efficient detection would then require separating the two light fields and using a local oscillator with the appropriate phase.

Instead, the information for the other degrees of freedom (DOF), especially the DOF perpendicular to the polarization axis, is scattered mainly perpendicularly to the beam propagation~\cite{maurer2023quantum}. 
In our case, considering a SW along $y$ polarized along $z$, the information about the x-motion is scattered mainly the fibers along $x$ (see SM-IV). This also applies to the $y$-motion using the fiber along $y$. Hence, to detect the in-plane motion, we collect the scattered light at $\lambda_y= 1550\mathrm{nm}$ ($\lambda_x= 1064\mathrm{nm}$) with the fibers along $x$ ($y$) and use it in a balanced homodyne scheme to detect the motion along $x$ ($y$). For detecting the motion along the $z$ direction, we use the second fiber along $y$. Here the multimode features of the SMF-28 fiber at $\lambda_x=1064\text{nm}$ allow the excitation of higher order modes, increasing the sensitivity to displacements along $z$. 

\begin{figure}
    \centering
    \includegraphics[width=1\linewidth]{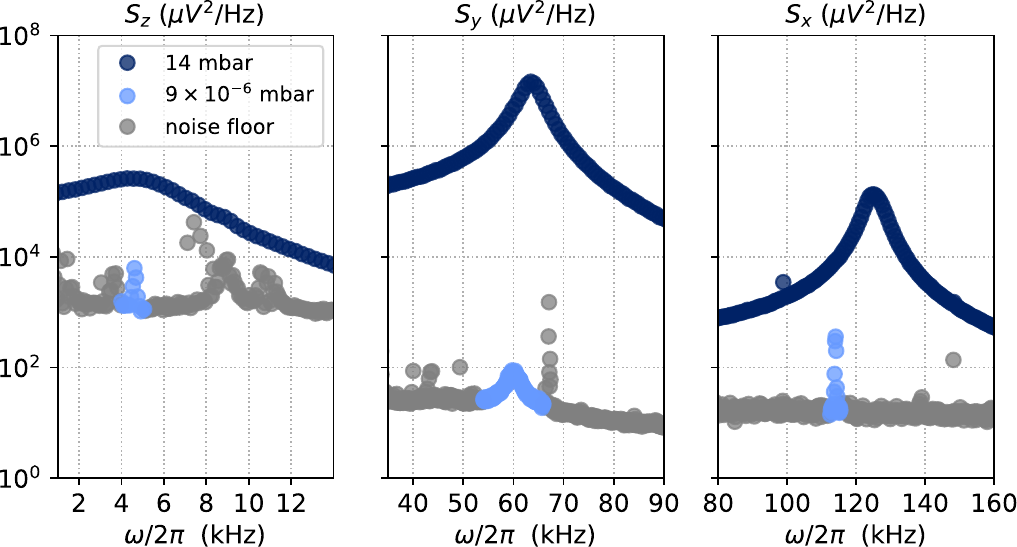}
    \caption{\textbf{PSDs of the particle's motion.} From left to right: PSD of the motion along $z$, $y$ and $x$, showing resonance frequencies of $\Omega_{x,y,z}/(2\pi) \approx(120,65,5)$ kHz. At $p=$14 mbar (dark blue), no feedback cooling is applied. At $p=9\times 10^{-6}$ mbar (light blue), active feedback stabilizes the particle's motion such that the area of the PSD is decreased. At $\omega\neq\Omega_j$ (grey), the detection noise dominates.}
    \label{fig:3}
\end{figure}

The power spectral densities (PSD) of the motion along $x,y,z$ at a pressure $p = 14$mbar are plotted in Fig~\ref{fig:3} (dark blue).  We reach an optimal SNR$_y\approx10^{6}$ comparable to forward scattering detection with high NA optics in standard experiments~\cite{gieseler_thesis}.


To cool and stabilize the nanoparticle's COM motion in vacuum, we apply electrical cold damping~\cite{Conangla_OptimalFB, Tebbenjohanns_ColdDamp} along $q=x,y,z$. The nanoparticle's COM motion is modeled as three decoupled harmonic oscillators described by the equation of motion

\begin{equation}
    \ddot{q}(t) + \Gamma_{m} \Dot{q}(t) + \Omega_q^2q(t) = \frac{F_{th}(t) + F_q^{\text{fb}}(t)}{m},
\end{equation}

where $m$ is the oscillator's mass~\cite{hebestreit_calibration_2018}, $\Gamma_m$ is the mechanical damping rate due to the surrounding gas, leading to a stochastic force $F_{th}(t)= \sigma \eta(t)= \sqrt{2k_B T m \Gamma_m }\eta(t)$ with $\eta(t)$ being white noise with unit standard deviation and zero mean~\cite{kubo1966fluctuation}. The feedback force is proportional to the delayed position, i.e., $F_q^{\text{fb}}(t)\propto k_q^d\: {q}(t-\tau)$ where $k_q^d$ is an adjustable gain and $\tau$ is a tunable delay. For $\tau=\pi/(2\Omega_q)$ this leads to an effective damping rate $\Gamma^{\text{eff}}_q=\Gamma_m + \Gamma^{\text{fb}}_q$ optimized for cooling~\cite{Conangla_OptimalFB}. 

The externally applied feedback force $F_q^{\text{fb}}(t) = Q E_q(t)$ depends on the charge of the particle $Q= n_q e^-$ with $n_q$ elementary charges and the homogeneous electric field $E_q(t) \propto \pm V_q(t)$ generated by planar electrodes as shown in Fig.~\ref{fig:1}a. For the in-plane DOF $x$ $(y)$, we apply two voltage signals $\pm V_x(t)$ ($\pm V_y(t)$) of equal amplitude but out of phase to the pair of electrodes situated left and right (top and bottom) of the chip. For cooling the $z$-direction, we apply $V_{z}(t)$ to a single planar electrode depicted in the centre of Fig.~\ref{fig:1}a. The symmetric electrode layout and the electrode-particle distance ensure the homogeneity of $E_q(t)$ (see SM-III). 

The results of 3D cold damping on-chip at vacuum are shown in Fig.~\ref{fig:3}. The individual panels display the Lorentzian PSDs of the particle displacement $S_q^{\text{IL}}(\omega)$ at pressures $p=14\text{mbar}$ (dark blue) and $p=9\times 10^{-6}\text{mbar}$ (light blue). In the latter, we stabilize the particle using active feedback in 3D. 

The lowest temperature $T_q^{\text{eff}}$ achievable in cold damping is determined by the detection noise $\sigma_q$ and the mechanical damping $\Gamma_m \propto p$~\cite{li2011millikelvin}. We fit the PSD of the in loop (IL) detector~\cite{Conangla_OptimalFB}

\begin{eqnarray}\label{eq:SIL}
      S_q^{\text{IL}}(\omega)   = \frac{\sigma^2/m^2}{(\Omega_q^2 -\omega^2)^2 + (\Gamma_m + \Gamma_q^{\text{fb}})^2\omega^2} + \\ \nonumber 
    \frac{(\Omega_q^2 - \omega^2)^2 + \Gamma_m^2\omega^2}{(\Omega_q^2 -\omega^2)^2 + (\Gamma_m + \Gamma_q^{\text{fb}})^2\omega^2} \sigma_q^{2}
\end{eqnarray}

to our data (see SM-V). This allows us to determine the oscillator's effective COM temperature $T_q^{\text{eff}}$ as 

\begin{equation}\label{eq:T_eff}
   T_q^{\text{eff}} = \frac{m \Omega_q^2}{2k_B} \left( \frac{\sigma^2/m^2}{\Omega_q^2(\Gamma_m + \Gamma_q^{\text{fb}})} + \frac{(\Gamma_q^{\text{fb}}\sigma_q)^2}{\Gamma_m + \Gamma_q^{\text{fb}}}\right),
\end{equation}
 where $ \Gamma_q^{\text{fb}} \propto\ k_q^d$. The phonon occupation then is 
 \begin{equation}\label{eq:Teff2phonon}
     \langle n_q\rangle=  \frac{k_B T_q^{\text{eff}}}{\hbar\Omega_q}
 \end{equation}

\begin{figure}
    \centering
    \includegraphics[width=\columnwidth]{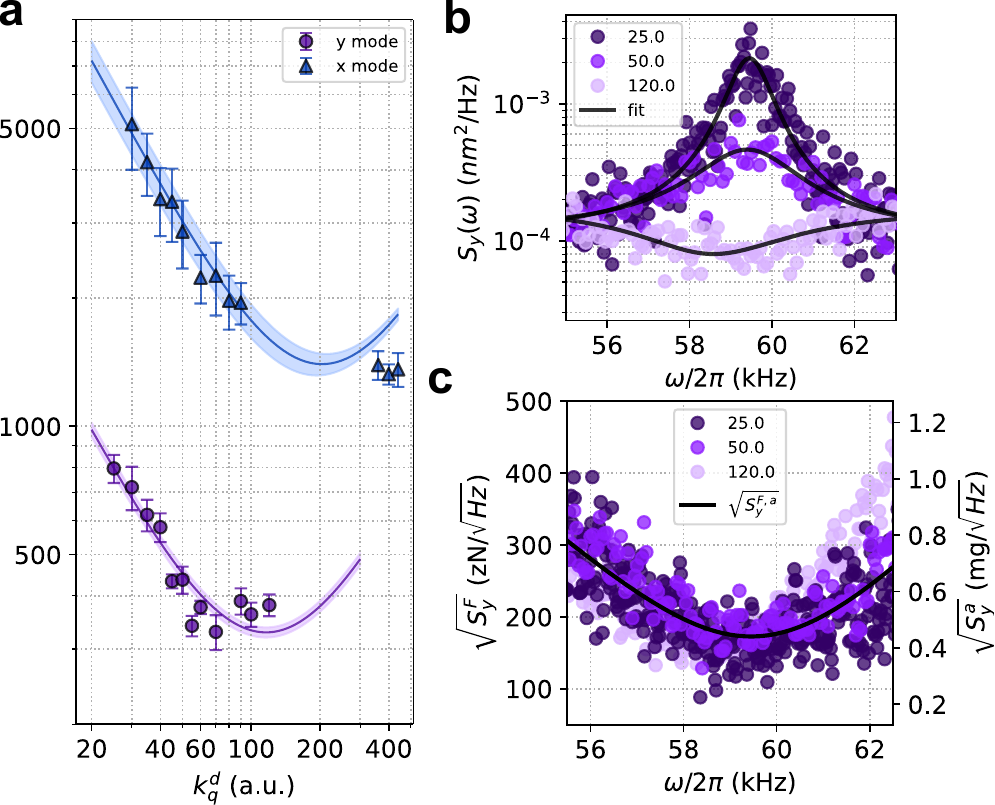}
    \caption{\textbf{Feedback cooling and sensitivity.} \textbf{a)} Mean phonon occupation of the motion $\langle n_q \rangle$along $x$ (blue triangles) and $y$ (purple circles) for increasing feedback gains $k_q^d$, extracted from the IL PSDs. The error bars are the standard deviation of 10 independent temperature measurements. The solid lines are fits to Eq. \ref{eq:Teff2phonon} with $T_q^{\text{eff}}$ given by Eq. \ref{eq:T_eff} where $\Gamma_q^{\text{fb}}$ is the only free parameter. Other parameters are extracted from the IL PSDs. The shaded regions correspond to the uncertainty of the extracted parameters (see SM-VI). \textbf{b)} PSD of the $y$ -motion for three different values of the feedback gain. The solid lines are fits to Eq. \ref{eq:SIL} (see text for details). \textbf{c)} PSDs of the force and acceleration sensitivities for the same three feedback gains as in \textbf{b}. The solid line is the minimum force and acceleration sensitivity from Eq. \ref{eq:SF} and Eq.~\ref{eq:Sa}.}
    \label{fig:4}
\end{figure}
 
Fig.~\ref{fig:4}a displays the estimated phonon occupation $\langle n_q\rangle$ for increasing feedback gain $k_q^d$ along the $x$ (blue triangle) and $y$ (purple circles) directions. The phonon occupation decreases to $\langle n_y\rangle =329\pm30$ and $\langle n_x\rangle = 1325\pm72$, respectively. The solid lines represent fits to Eq.~\ref{eq:Teff2phonon} with the shaded regions indicating the uncertainty associated with the fitted curve (see SM-VI). For large feedback gains, $\langle n_q\rangle$ increases due to correlations between the detector noise $\sigma_q$ and feedback signal, as depicted in Fig.~\ref{fig:4}a. In the PSD, it manifests as the so-called noise squashing~\cite{poggio2007feedback,Conangla_OptimalFB,Tebbenjohanns_ColdDamp}, such that $S_q^{\text{IL}}(\Omega_q)$ yields non-physical values below the noise floor. This is depicted in Fig.~\ref{fig:4}b, where we plot the measured PSD of the $y$ motion for different feedback gain values. Note that we use the in-loop detection signal for the temperature estimation. This approach remains valid, given that we account for the noise squashing by adding the second term in Eq.~\ref{eq:T_eff} ~\cite{Conangla_OptimalFB}, making an out-of-loop detector unnecessary for the temperature estimation.
    

Due to its compact design and excellent cooling performance, the developed platform shows promises for application in force and acceleration sensing. The minimum force and acceleration sensitivity taking into account detection noise (see SM-VII) exceeds the thermal limit and is given by 

\begin{align}\label{eq:SF}
    \sqrt{S_q^{F}} &= \sqrt{{2 m k_B T\Gamma_m}+m^2[(\Omega_q^2-\omega^2)^2+\omega^2\Gamma_m^2]\sigma_q^2}\\
    \label{eq:Sa}
    \sqrt{S_q^{a}} &= \sqrt{2 k_B T \Gamma_m/m+[(\Omega_q^2-\omega^2)^2+\omega^2\Gamma_m^2]\sigma_q^2}
\end{align}

and is  independent from the cooling of the COM. Yet, for real-life applications, the detection noise $\sigma_q$ limits the bandwidth in which minimal sensitivities are achievable. 

In Fig.~\ref{fig:4}c, we display the force and acceleration spectral densities, $\sqrt{S_y^{F}}$ and $\sqrt{S_y^{a}}$, for a range of $k_y^d=[25,50,120]$. The solid line corresponds to Eq.~\ref{eq:SF} and ~\ref{eq:Sa} using the parameters extracted from the fitted IL PSDs. The experimentally achieved minimum sensitivities are $\sqrt{S_y^{F}}= (176\pm 26) \text{zN}/ \sqrt{\text{Hz}}$ and $\sqrt{S_y^{a}}=(440\pm70) \mu\text{g} / \sqrt{\text{Hz}}$ which increase by 3dB over a 8kHz bandwidth around $\Omega_y/(2\pi)= 59 \text{kHz}$ due to detector noise contributions exceeding the thermal noise. As expected, both bandwidth and sensitivities show no dependence of $k_y^d$. Note that the hypothetical additional use of an out-of-loop detector would further deteriorate the sensitivities since we would have noise contributions from both detectors (see SM-VII). 


In conclusion, we have demonstrated robust optical on-chip levitation and motion control in vacuum with a fully integrated platform. Despite the use of commercial fibers with low NA, we have shown that the particle's displacement detection reaches comparable signal-to-noise ratios as with bulky high NA optics. Such performance, that enables already the cooling of the COM down to hundreds of phonons, can be further enhanced by further reducing the fiber distance.

The implemented platform also offers interesting applications towards multiparticle arrays~\cite{vijayan2023scalable} beyond ten particles, optical binding~\cite{rieser2022tunable,burns1989optical} and the levitation of high refractive index meta-atoms~\cite{lepeshov2023levitated}. Additionally, as previously mentioned, our fabrication technique also enables more complex lattice forms~\cite{windpassinger2013engineering, li2021non} by controlled in-plane fiber positioning.

Furthermore, our micro-fabricated fiber mounts are suitable for the integration of particle loading mechanisms based on hollow core fibers~\cite{grass2016optical,bykov2018long}, which up to today is the only technique that offers direct particle loading into optical traps in ultra-high vacuum conditions, which is an important aspect for the generation of macroscopic quantum superposition states~\cite{roda2023macroscopic,stickler2018probing}.

In the future, our on-chip platform can benefit from the use of micro-fabricated optics~\cite{gonzalez2023micro} or meta surfaces on fibers~\cite{plidschun2021ultrahigh,ren2022achromatic} for better detection and trapping. Furthermore, more complex optical elements like fiber cavities~\cite{steinmetz2006stable,hunger2010fiber,muller2010ultrahigh} can also be integrated.  We envision our platform as the initial stepping stone towards the use of hybrid potentials for quantum experiments based on levitated particles. 


\begin{acknowledgments}
The authors acknowledge financial support from the European Commission through grant IQLev (grant Agreement No. 863132) as well as the European Research Council through grant Q-Xtreme ERC 2020-SyG (grant Agreement No. 951234)
\end{acknowledgments}


\begin{thebibliography}{63}%
	\makeatletter
	\providecommand \@ifxundefined [1]{%
		\@ifx{#1\undefined}
	}%
	\providecommand \@ifnum [1]{%
		\ifnum #1\expandafter \@firstoftwo
		\else \expandafter \@secondoftwo
		\fi
	}%
	\providecommand \@ifx [1]{%
		\ifx #1\expandafter \@firstoftwo
		\else \expandafter \@secondoftwo
		\fi
	}%
	\providecommand \natexlab [1]{#1}%
	\providecommand \enquote  [1]{``#1''}%
	\providecommand \bibnamefont  [1]{#1}%
	\providecommand \bibfnamefont [1]{#1}%
	\providecommand \citenamefont [1]{#1}%
	\providecommand \href@noop [0]{\@secondoftwo}%
	\providecommand \href [0]{\begingroup \@sanitize@url \@href}%
	\providecommand \@href[1]{\@@startlink{#1}\@@href}%
	\providecommand \@@href[1]{\endgroup#1\@@endlink}%
	\providecommand \@sanitize@url [0]{\catcode `\\12\catcode `\$12\catcode
		`\&12\catcode `\#12\catcode `\^12\catcode `\_12\catcode `\%12\relax}%
	\providecommand \@@startlink[1]{}%
	\providecommand \@@endlink[0]{}%
	\providecommand \url  [0]{\begingroup\@sanitize@url \@url }%
	\providecommand \@url [1]{\endgroup\@href {#1}{\urlprefix }}%
	\providecommand \urlprefix  [0]{URL }%
	\providecommand \Eprint [0]{\href }%
	\providecommand \doibase [0]{https://doi.org/}%
	\providecommand \selectlanguage [0]{\@gobble}%
	\providecommand \bibinfo  [0]{\@secondoftwo}%
	\providecommand \bibfield  [0]{\@secondoftwo}%
	\providecommand \translation [1]{[#1]}%
	\providecommand \BibitemOpen [0]{}%
	\providecommand \bibitemStop [0]{}%
	\providecommand \bibitemNoStop [0]{.\EOS\space}%
	\providecommand \EOS [0]{\spacefactor3000\relax}%
	\providecommand \BibitemShut  [1]{\csname bibitem#1\endcsname}%
	\let\auto@bib@innerbib\@empty
	\bibitem [{\citenamefont {Ashkin}\ and\ \citenamefont
		{Dziedzic}(1971)}]{ashkin1971optical}%
	\BibitemOpen
	\bibfield  {author} {\bibinfo {author} {\bibfnamefont {A.}~\bibnamefont
			{Ashkin}}\ and\ \bibinfo {author} {\bibfnamefont {J.}~\bibnamefont
			{Dziedzic}},\ }\href@noop {} {\bibfield  {journal} {\bibinfo  {journal}
			{Applied Physics Letters}\ }\textbf {\bibinfo {volume} {19}},\ \bibinfo
		{pages} {283} (\bibinfo {year} {1971})}\BibitemShut {NoStop}%
	\bibitem [{\citenamefont {Gonzalez-Ballestero}\ \emph
		{et~al.}(2021)\citenamefont {Gonzalez-Ballestero}, \citenamefont
		{Aspelmeyer}, \citenamefont {Novotny}, \citenamefont {Quidant},\ and\
		\citenamefont {Romero-Isart}}]{gonzalez2021levitodynamics}%
	\BibitemOpen
	\bibfield  {author} {\bibinfo {author} {\bibfnamefont {C.}~\bibnamefont
			{Gonzalez-Ballestero}}, \bibinfo {author} {\bibfnamefont {M.}~\bibnamefont
			{Aspelmeyer}}, \bibinfo {author} {\bibfnamefont {L.}~\bibnamefont {Novotny}},
		\bibinfo {author} {\bibfnamefont {R.}~\bibnamefont {Quidant}},\ and\ \bibinfo
		{author} {\bibfnamefont {O.}~\bibnamefont {Romero-Isart}},\ }\href@noop {}
	{\bibfield  {journal} {\bibinfo  {journal} {Science}\ }\textbf {\bibinfo
			{volume} {374}},\ \bibinfo {pages} {eabg3027} (\bibinfo {year}
		{2021})}\BibitemShut {NoStop}%
	\bibitem [{\citenamefont {Jain}\ \emph {et~al.}(2016)\citenamefont {Jain},
		\citenamefont {Gieseler}, \citenamefont {Moritz}, \citenamefont {Dellago},
		\citenamefont {Quidant},\ and\ \citenamefont {Novotny}}]{jain_direct_2016}%
	\BibitemOpen
	\bibfield  {author} {\bibinfo {author} {\bibfnamefont {V.}~\bibnamefont
			{Jain}}, \bibinfo {author} {\bibfnamefont {J.}~\bibnamefont {Gieseler}},
		\bibinfo {author} {\bibfnamefont {C.}~\bibnamefont {Moritz}}, \bibinfo
		{author} {\bibfnamefont {C.}~\bibnamefont {Dellago}}, \bibinfo {author}
		{\bibfnamefont {R.}~\bibnamefont {Quidant}},\ and\ \bibinfo {author}
		{\bibfnamefont {L.}~\bibnamefont {Novotny}},\ }\href
	{https://doi.org/10.1103/PhysRevLett.116.243601} {\bibfield  {journal}
		{\bibinfo  {journal} {Phys. Rev. Lett.}\ }\textbf {\bibinfo {volume} {116}},\
		\bibinfo {pages} {243601} (\bibinfo {year} {2016})}\BibitemShut {NoStop}%
	\bibitem [{\citenamefont {Millen}\ \emph {et~al.}(2015)\citenamefont {Millen},
		\citenamefont {Fonseca}, \citenamefont {Mavrogordatos}, \citenamefont
		{Monteiro},\ and\ \citenamefont {Barker}}]{millen2015cavity}%
	\BibitemOpen
	\bibfield  {author} {\bibinfo {author} {\bibfnamefont {J.}~\bibnamefont
			{Millen}}, \bibinfo {author} {\bibfnamefont {P.}~\bibnamefont {Fonseca}},
		\bibinfo {author} {\bibfnamefont {T.}~\bibnamefont {Mavrogordatos}}, \bibinfo
		{author} {\bibfnamefont {T.}~\bibnamefont {Monteiro}},\ and\ \bibinfo
		{author} {\bibfnamefont {P.}~\bibnamefont {Barker}},\ }\href@noop {}
	{\bibfield  {journal} {\bibinfo  {journal} {Physical review letters}\
		}\textbf {\bibinfo {volume} {114}},\ \bibinfo {pages} {123602} (\bibinfo
		{year} {2015})}\BibitemShut {NoStop}%
	\bibitem [{\citenamefont {Conangla}\ \emph {et~al.}(2020)\citenamefont
		{Conangla}, \citenamefont {Rica},\ and\ \citenamefont
		{Quidant}}]{conangla2020extending}%
	\BibitemOpen
	\bibfield  {author} {\bibinfo {author} {\bibfnamefont {G.~P.}\ \bibnamefont
			{Conangla}}, \bibinfo {author} {\bibfnamefont {R.~A.}\ \bibnamefont {Rica}},\
		and\ \bibinfo {author} {\bibfnamefont {R.}~\bibnamefont {Quidant}},\
	}\href@noop {} {\bibfield  {journal} {\bibinfo  {journal} {Nano Letters}\
		}\textbf {\bibinfo {volume} {20}},\ \bibinfo {pages} {6018} (\bibinfo {year}
		{2020})}\BibitemShut {NoStop}%
	\bibitem [{\citenamefont {Bykov}\ \emph {et~al.}(2022)\citenamefont {Bykov},
		\citenamefont {Meusburger}, \citenamefont {Dania},\ and\ \citenamefont
		{Northup}}]{bykov2022hybrid}%
	\BibitemOpen
	\bibfield  {author} {\bibinfo {author} {\bibfnamefont {D.~S.}\ \bibnamefont
			{Bykov}}, \bibinfo {author} {\bibfnamefont {M.}~\bibnamefont {Meusburger}},
		\bibinfo {author} {\bibfnamefont {L.}~\bibnamefont {Dania}},\ and\ \bibinfo
		{author} {\bibfnamefont {T.~E.}\ \bibnamefont {Northup}},\ }\href@noop {}
	{\bibfield  {journal} {\bibinfo  {journal} {Review of Scientific
				Instruments}\ }\textbf {\bibinfo {volume} {93}},\ \bibinfo {pages} {073201}
		(\bibinfo {year} {2022})}\BibitemShut {NoStop}%
	\bibitem [{\citenamefont {Conangla}\ \emph {et~al.}(2019)\citenamefont
		{Conangla}, \citenamefont {Ricci}, \citenamefont {Cuairan}, \citenamefont
		{Schell}, \citenamefont {Meyer},\ and\ \citenamefont
		{Quidant}}]{Conangla_OptimalFB}%
	\BibitemOpen
	\bibfield  {author} {\bibinfo {author} {\bibfnamefont {G.~P.}\ \bibnamefont
			{Conangla}}, \bibinfo {author} {\bibfnamefont {F.}~\bibnamefont {Ricci}},
		\bibinfo {author} {\bibfnamefont {M.~T.}\ \bibnamefont {Cuairan}}, \bibinfo
		{author} {\bibfnamefont {A.~W.}\ \bibnamefont {Schell}}, \bibinfo {author}
		{\bibfnamefont {N.}~\bibnamefont {Meyer}},\ and\ \bibinfo {author}
		{\bibfnamefont {R.}~\bibnamefont {Quidant}},\ }\href
	{https://doi.org/10.1103/PhysRevLett.122.223602} {\bibfield  {journal}
		{\bibinfo  {journal} {Phys. Rev. Lett.}\ }\textbf {\bibinfo {volume} {122}},\
		\bibinfo {pages} {223602} (\bibinfo {year} {2019})}\BibitemShut {NoStop}%
	\bibitem [{\citenamefont {Tebbenjohanns}\ \emph
		{et~al.}(2019{\natexlab{a}})\citenamefont {Tebbenjohanns}, \citenamefont
		{Frimmer}, \citenamefont {Militaru}, \citenamefont {Jain},\ and\
		\citenamefont {Novotny}}]{Tebbenjohanns_ColdDamp}%
	\BibitemOpen
	\bibfield  {author} {\bibinfo {author} {\bibfnamefont {F.}~\bibnamefont
			{Tebbenjohanns}}, \bibinfo {author} {\bibfnamefont {M.}~\bibnamefont
			{Frimmer}}, \bibinfo {author} {\bibfnamefont {A.}~\bibnamefont {Militaru}},
		\bibinfo {author} {\bibfnamefont {V.}~\bibnamefont {Jain}},\ and\ \bibinfo
		{author} {\bibfnamefont {L.}~\bibnamefont {Novotny}},\ }\href
	{https://doi.org/10.1103/PhysRevLett.122.223601} {\bibfield  {journal}
		{\bibinfo  {journal} {Phys. Rev. Lett.}\ }\textbf {\bibinfo {volume} {122}},\
		\bibinfo {pages} {223601} (\bibinfo {year} {2019}{\natexlab{a}})}\BibitemShut
	{NoStop}%
	\bibitem [{\citenamefont {Gieseler}\ \emph {et~al.}(2012)\citenamefont
		{Gieseler}, \citenamefont {Deutsch}, \citenamefont {Quidant},\ and\
		\citenamefont {Novotny}}]{gieseler_subkelvin_2012}%
	\BibitemOpen
	\bibfield  {author} {\bibinfo {author} {\bibfnamefont {J.}~\bibnamefont
			{Gieseler}}, \bibinfo {author} {\bibfnamefont {B.}~\bibnamefont {Deutsch}},
		\bibinfo {author} {\bibfnamefont {R.}~\bibnamefont {Quidant}},\ and\ \bibinfo
		{author} {\bibfnamefont {L.}~\bibnamefont {Novotny}},\ }\href
	{https://doi.org/10.1103/PhysRevLett.109.103603} {\bibfield  {journal}
		{\bibinfo  {journal} {Phys. Rev. Lett.}\ }\textbf {\bibinfo {volume} {109}},\
		\bibinfo {pages} {103603} (\bibinfo {year} {2012})}\BibitemShut {NoStop}%
	\bibitem [{\citenamefont {Timberlake}\ \emph {et~al.}(2019)\citenamefont
		{Timberlake}, \citenamefont {Gasbarri}, \citenamefont {Vinante},
		\citenamefont {Setter},\ and\ \citenamefont
		{Ulbricht}}]{timberlake2019acceleration}%
	\BibitemOpen
	\bibfield  {author} {\bibinfo {author} {\bibfnamefont {C.}~\bibnamefont
			{Timberlake}}, \bibinfo {author} {\bibfnamefont {G.}~\bibnamefont
			{Gasbarri}}, \bibinfo {author} {\bibfnamefont {A.}~\bibnamefont {Vinante}},
		\bibinfo {author} {\bibfnamefont {A.}~\bibnamefont {Setter}},\ and\ \bibinfo
		{author} {\bibfnamefont {H.}~\bibnamefont {Ulbricht}},\ }\href@noop {}
	{\bibfield  {journal} {\bibinfo  {journal} {Applied Physics Letters}\
		}\textbf {\bibinfo {volume} {115}} (\bibinfo {year} {2019})}\BibitemShut
	{NoStop}%
	\bibitem [{\citenamefont {Monteiro}\ \emph {et~al.}(2020)\citenamefont
		{Monteiro}, \citenamefont {Li}, \citenamefont {Afek}, \citenamefont {Li},
		\citenamefont {Mossman},\ and\ \citenamefont
		{Moore}}]{Monteiro_force_acceleration_sensing_2020}%
	\BibitemOpen
	\bibfield  {author} {\bibinfo {author} {\bibfnamefont {F.}~\bibnamefont
			{Monteiro}}, \bibinfo {author} {\bibfnamefont {W.}~\bibnamefont {Li}},
		\bibinfo {author} {\bibfnamefont {G.}~\bibnamefont {Afek}}, \bibinfo {author}
		{\bibfnamefont {C.-l.}\ \bibnamefont {Li}}, \bibinfo {author} {\bibfnamefont
			{M.}~\bibnamefont {Mossman}},\ and\ \bibinfo {author} {\bibfnamefont {D.~C.}\
			\bibnamefont {Moore}},\ }\href {https://doi.org/10.1103/PhysRevA.101.053835}
	{\bibfield  {journal} {\bibinfo  {journal} {Phys. Rev. A}\ }\textbf {\bibinfo
			{volume} {101}},\ \bibinfo {pages} {053835} (\bibinfo {year}
		{2020})}\BibitemShut {NoStop}%
	\bibitem [{\citenamefont {Winstone}\ \emph {et~al.}(2018)\citenamefont
		{Winstone}, \citenamefont {Bennett}, \citenamefont {Rademacher},
		\citenamefont {Rashid}, \citenamefont {Buhmann},\ and\ \citenamefont
		{Ulbricht}}]{winstone_electrostaticimage_force_sensing}%
	\BibitemOpen
	\bibfield  {author} {\bibinfo {author} {\bibfnamefont {G.}~\bibnamefont
			{Winstone}}, \bibinfo {author} {\bibfnamefont {R.}~\bibnamefont {Bennett}},
		\bibinfo {author} {\bibfnamefont {M.}~\bibnamefont {Rademacher}}, \bibinfo
		{author} {\bibfnamefont {M.}~\bibnamefont {Rashid}}, \bibinfo {author}
		{\bibfnamefont {S.}~\bibnamefont {Buhmann}},\ and\ \bibinfo {author}
		{\bibfnamefont {H.}~\bibnamefont {Ulbricht}},\ }\href
	{https://doi.org/10.1103/PhysRevA.98.053831} {\bibfield  {journal} {\bibinfo
			{journal} {Phys. Rev. A}\ }\textbf {\bibinfo {volume} {98}},\ \bibinfo
		{pages} {053831} (\bibinfo {year} {2018})}\BibitemShut {NoStop}%
	\bibitem [{\citenamefont {Hempston}\ \emph {et~al.}(2017)\citenamefont
		{Hempston}, \citenamefont {Vovrosh}, \citenamefont {Toroš}, \citenamefont
		{Winstone}, \citenamefont {Rashid},\ and\ \citenamefont
		{Ulbricht}}]{Hempston_force}%
	\BibitemOpen
	\bibfield  {author} {\bibinfo {author} {\bibfnamefont {D.}~\bibnamefont
			{Hempston}}, \bibinfo {author} {\bibfnamefont {J.}~\bibnamefont {Vovrosh}},
		\bibinfo {author} {\bibfnamefont {M.}~\bibnamefont {Toroš}}, \bibinfo
		{author} {\bibfnamefont {G.}~\bibnamefont {Winstone}}, \bibinfo {author}
		{\bibfnamefont {M.}~\bibnamefont {Rashid}},\ and\ \bibinfo {author}
		{\bibfnamefont {H.}~\bibnamefont {Ulbricht}},\ }\href
	{https://doi.org/10.1063/1.4993555} {\bibfield  {journal} {\bibinfo
			{journal} {Applied Physics Letters}\ }\textbf {\bibinfo {volume} {111}},\
		\bibinfo {pages} {133111} (\bibinfo {year} {2017})}\BibitemShut {NoStop}%
	\bibitem [{\citenamefont {Ranjit}\ \emph {et~al.}(2016)\citenamefont {Ranjit},
		\citenamefont {Cunningham}, \citenamefont {Casey},\ and\ \citenamefont
		{Geraci}}]{ranjit2016zeptonewton}%
	\BibitemOpen
	\bibfield  {author} {\bibinfo {author} {\bibfnamefont {G.}~\bibnamefont
			{Ranjit}}, \bibinfo {author} {\bibfnamefont {M.}~\bibnamefont {Cunningham}},
		\bibinfo {author} {\bibfnamefont {K.}~\bibnamefont {Casey}},\ and\ \bibinfo
		{author} {\bibfnamefont {A.~A.}\ \bibnamefont {Geraci}},\ }\href@noop {}
	{\bibfield  {journal} {\bibinfo  {journal} {Physical Review A}\ }\textbf
		{\bibinfo {volume} {93}},\ \bibinfo {pages} {053801} (\bibinfo {year}
		{2016})}\BibitemShut {NoStop}%
	\bibitem [{\citenamefont {Kawasaki}\ \emph {et~al.}(2020)\citenamefont
		{Kawasaki}, \citenamefont {Fieguth}, \citenamefont {Priel}, \citenamefont
		{Blakemore}, \citenamefont {Martin},\ and\ \citenamefont
		{Gratta}}]{Kawasaki_short-distance_2020}%
	\BibitemOpen
	\bibfield  {author} {\bibinfo {author} {\bibfnamefont {A.}~\bibnamefont
			{Kawasaki}}, \bibinfo {author} {\bibfnamefont {A.}~\bibnamefont {Fieguth}},
		\bibinfo {author} {\bibfnamefont {N.}~\bibnamefont {Priel}}, \bibinfo
		{author} {\bibfnamefont {C.~P.}\ \bibnamefont {Blakemore}}, \bibinfo {author}
		{\bibfnamefont {D.}~\bibnamefont {Martin}},\ and\ \bibinfo {author}
		{\bibfnamefont {G.}~\bibnamefont {Gratta}},\ }\href
	{https://doi.org/10.1063/5.0011759} {\bibfield  {journal} {\bibinfo
			{journal} {Review of Scientific Instruments}\ }\textbf {\bibinfo {volume}
			{91}},\ \bibinfo {pages} {083201} (\bibinfo {year} {2020})}\BibitemShut
	{NoStop}%
	\bibitem [{\citenamefont {Hebestreit}\ \emph
		{et~al.}(2018{\natexlab{a}})\citenamefont {Hebestreit}, \citenamefont
		{Frimmer}, \citenamefont {Reimann},\ and\ \citenamefont
		{Novotny}}]{Hebestreit_staticforce}%
	\BibitemOpen
	\bibfield  {author} {\bibinfo {author} {\bibfnamefont {E.}~\bibnamefont
			{Hebestreit}}, \bibinfo {author} {\bibfnamefont {M.}~\bibnamefont {Frimmer}},
		\bibinfo {author} {\bibfnamefont {R.}~\bibnamefont {Reimann}},\ and\ \bibinfo
		{author} {\bibfnamefont {L.}~\bibnamefont {Novotny}},\ }\href
	{https://doi.org/10.1103/PhysRevLett.121.063602} {\bibfield  {journal}
		{\bibinfo  {journal} {Phys. Rev. Lett.}\ }\textbf {\bibinfo {volume} {121}},\
		\bibinfo {pages} {063602} (\bibinfo {year} {2018}{\natexlab{a}})}\BibitemShut
	{NoStop}%
	\bibitem [{\citenamefont {Ahn}\ \emph {et~al.}(2020)\citenamefont {Ahn},
		\citenamefont {Xu}, \citenamefont {Bang}, \citenamefont {Ju}, \citenamefont
		{Gao},\ and\ \citenamefont {Li}}]{ahn2020ultrasensitive}%
	\BibitemOpen
	\bibfield  {author} {\bibinfo {author} {\bibfnamefont {J.}~\bibnamefont
			{Ahn}}, \bibinfo {author} {\bibfnamefont {Z.}~\bibnamefont {Xu}}, \bibinfo
		{author} {\bibfnamefont {J.}~\bibnamefont {Bang}}, \bibinfo {author}
		{\bibfnamefont {P.}~\bibnamefont {Ju}}, \bibinfo {author} {\bibfnamefont
			{X.}~\bibnamefont {Gao}},\ and\ \bibinfo {author} {\bibfnamefont
			{T.}~\bibnamefont {Li}},\ }\href@noop {} {\bibfield  {journal} {\bibinfo
			{journal} {Nature Nanotechnology}\ }\textbf {\bibinfo {volume} {15}},\
		\bibinfo {pages} {89} (\bibinfo {year} {2020})}\BibitemShut {NoStop}%
	\bibitem [{\citenamefont {Magrini}\ \emph {et~al.}(2021)\citenamefont
		{Magrini}, \citenamefont {Rosenzweig}, \citenamefont {Bach}, \citenamefont
		{Deutschmann-Olek}, \citenamefont {Hofer}, \citenamefont {Hong},
		\citenamefont {Kiesel}, \citenamefont {Kugi},\ and\ \citenamefont
		{Aspelmeyer}}]{magrini_real-time_2021}%
	\BibitemOpen
	\bibfield  {author} {\bibinfo {author} {\bibfnamefont {L.}~\bibnamefont
			{Magrini}}, \bibinfo {author} {\bibfnamefont {P.}~\bibnamefont {Rosenzweig}},
		\bibinfo {author} {\bibfnamefont {C.}~\bibnamefont {Bach}}, \bibinfo {author}
		{\bibfnamefont {A.}~\bibnamefont {Deutschmann-Olek}}, \bibinfo {author}
		{\bibfnamefont {S.~G.}\ \bibnamefont {Hofer}}, \bibinfo {author}
		{\bibfnamefont {S.}~\bibnamefont {Hong}}, \bibinfo {author} {\bibfnamefont
			{N.}~\bibnamefont {Kiesel}}, \bibinfo {author} {\bibfnamefont
			{A.}~\bibnamefont {Kugi}},\ and\ \bibinfo {author} {\bibfnamefont
			{M.}~\bibnamefont {Aspelmeyer}},\ }\href
	{https://doi.org/10.1038/s41586-021-03602-3} {\bibfield  {journal} {\bibinfo
			{journal} {Nature}\ }\textbf {\bibinfo {volume} {595}},\ \bibinfo {pages}
		{373} (\bibinfo {year} {2021})}\BibitemShut {NoStop}%
	\bibitem [{\citenamefont {Tebbenjohanns}\ \emph {et~al.}(2021)\citenamefont
		{Tebbenjohanns}, \citenamefont {Mattana}, \citenamefont {Rossi},
		\citenamefont {Frimmer},\ and\ \citenamefont
		{Novotny}}]{tebbenjohanns_quantum_2021}%
	\BibitemOpen
	\bibfield  {author} {\bibinfo {author} {\bibfnamefont {F.}~\bibnamefont
			{Tebbenjohanns}}, \bibinfo {author} {\bibfnamefont {M.~L.}\ \bibnamefont
			{Mattana}}, \bibinfo {author} {\bibfnamefont {M.}~\bibnamefont {Rossi}},
		\bibinfo {author} {\bibfnamefont {M.}~\bibnamefont {Frimmer}},\ and\ \bibinfo
		{author} {\bibfnamefont {L.}~\bibnamefont {Novotny}},\ }\bibfield  {journal}
	{\bibinfo  {journal} {Nature}\ }\textbf {\bibinfo {volume} {595}},\ \href
	{https://doi.org/10.1038/s41586-021-03617-w} {10.1038/s41586-021-03617-w}
	(\bibinfo {year} {2021})\BibitemShut {NoStop}%
	\bibitem [{\citenamefont {Deli{\'c}}\ \emph {et~al.}(2020)\citenamefont
		{Deli{\'c}}, \citenamefont {Reisenbauer}, \citenamefont {Dare}, \citenamefont
		{Grass}, \citenamefont {Vuleti{\'c}}, \citenamefont {Kiesel},\ and\
		\citenamefont {Aspelmeyer}}]{delic2020cooling}%
	\BibitemOpen
	\bibfield  {author} {\bibinfo {author} {\bibfnamefont {U.}~\bibnamefont
			{Deli{\'c}}}, \bibinfo {author} {\bibfnamefont {M.}~\bibnamefont
			{Reisenbauer}}, \bibinfo {author} {\bibfnamefont {K.}~\bibnamefont {Dare}},
		\bibinfo {author} {\bibfnamefont {D.}~\bibnamefont {Grass}}, \bibinfo
		{author} {\bibfnamefont {V.}~\bibnamefont {Vuleti{\'c}}}, \bibinfo {author}
		{\bibfnamefont {N.}~\bibnamefont {Kiesel}},\ and\ \bibinfo {author}
		{\bibfnamefont {M.}~\bibnamefont {Aspelmeyer}},\ }\href@noop {} {\bibfield
		{journal} {\bibinfo  {journal} {Science}\ }\textbf {\bibinfo {volume}
			{367}},\ \bibinfo {pages} {892} (\bibinfo {year} {2020})}\BibitemShut
	{NoStop}%
	\bibitem [{\citenamefont {Ranfagni}\ \emph {et~al.}(2022)\citenamefont
		{Ranfagni}, \citenamefont {B{\o}rkje}, \citenamefont {Marino},\ and\
		\citenamefont {Marin}}]{ranfagni2022two}%
	\BibitemOpen
	\bibfield  {author} {\bibinfo {author} {\bibfnamefont {A.}~\bibnamefont
			{Ranfagni}}, \bibinfo {author} {\bibfnamefont {K.}~\bibnamefont {B{\o}rkje}},
		\bibinfo {author} {\bibfnamefont {F.}~\bibnamefont {Marino}},\ and\ \bibinfo
		{author} {\bibfnamefont {F.}~\bibnamefont {Marin}},\ }\href@noop {}
	{\bibfield  {journal} {\bibinfo  {journal} {Physical Review Research}\
		}\textbf {\bibinfo {volume} {4}},\ \bibinfo {pages} {033051} (\bibinfo {year}
		{2022})}\BibitemShut {NoStop}%
	\bibitem [{\citenamefont {Kamba}\ \emph {et~al.}(2022)\citenamefont {Kamba},
		\citenamefont {Shimizu},\ and\ \citenamefont {Aikawa}}]{kamba2022optical}%
	\BibitemOpen
	\bibfield  {author} {\bibinfo {author} {\bibfnamefont {M.}~\bibnamefont
			{Kamba}}, \bibinfo {author} {\bibfnamefont {R.}~\bibnamefont {Shimizu}},\
		and\ \bibinfo {author} {\bibfnamefont {K.}~\bibnamefont {Aikawa}},\
	}\href@noop {} {\bibfield  {journal} {\bibinfo  {journal} {Optics Express}\
		}\textbf {\bibinfo {volume} {30}},\ \bibinfo {pages} {26716} (\bibinfo {year}
		{2022})}\BibitemShut {NoStop}%
	\bibitem [{\citenamefont {Piotrowski}\ \emph {et~al.}(2023)\citenamefont
		{Piotrowski}, \citenamefont {Windey}, \citenamefont {Vijayan}, \citenamefont
		{Gonzalez-Ballestero}, \citenamefont {de~los R{\'\i}os~Sommer}, \citenamefont
		{Meyer}, \citenamefont {Quidant}, \citenamefont {Romero-Isart}, \citenamefont
		{Reimann},\ and\ \citenamefont {Novotny}}]{piotrowski2023simultaneous}%
	\BibitemOpen
	\bibfield  {author} {\bibinfo {author} {\bibfnamefont {J.}~\bibnamefont
			{Piotrowski}}, \bibinfo {author} {\bibfnamefont {D.}~\bibnamefont {Windey}},
		\bibinfo {author} {\bibfnamefont {J.}~\bibnamefont {Vijayan}}, \bibinfo
		{author} {\bibfnamefont {C.}~\bibnamefont {Gonzalez-Ballestero}}, \bibinfo
		{author} {\bibfnamefont {A.}~\bibnamefont {de~los R{\'\i}os~Sommer}},
		\bibinfo {author} {\bibfnamefont {N.}~\bibnamefont {Meyer}}, \bibinfo
		{author} {\bibfnamefont {R.}~\bibnamefont {Quidant}}, \bibinfo {author}
		{\bibfnamefont {O.}~\bibnamefont {Romero-Isart}}, \bibinfo {author}
		{\bibfnamefont {R.}~\bibnamefont {Reimann}},\ and\ \bibinfo {author}
		{\bibfnamefont {L.}~\bibnamefont {Novotny}},\ }\href@noop {} {\bibfield
		{journal} {\bibinfo  {journal} {Nature Physics}\ ,\ \bibinfo {pages} {1}}
		(\bibinfo {year} {2023})}\BibitemShut {NoStop}%
	\bibitem [{\citenamefont {Bruzewicz}\ \emph {et~al.}(2019)\citenamefont
		{Bruzewicz}, \citenamefont {Chiaverini}, \citenamefont {McConnell},\ and\
		\citenamefont {Sage}}]{bruzewicz2019trapped}%
	\BibitemOpen
	\bibfield  {author} {\bibinfo {author} {\bibfnamefont {C.~D.}\ \bibnamefont
			{Bruzewicz}}, \bibinfo {author} {\bibfnamefont {J.}~\bibnamefont
			{Chiaverini}}, \bibinfo {author} {\bibfnamefont {R.}~\bibnamefont
			{McConnell}},\ and\ \bibinfo {author} {\bibfnamefont {J.~M.}\ \bibnamefont
			{Sage}},\ }\href@noop {} {\bibfield  {journal} {\bibinfo  {journal} {Applied
				Physics Reviews}\ }\textbf {\bibinfo {volume} {6}} (\bibinfo {year}
		{2019})}\BibitemShut {NoStop}%
	\bibitem [{\citenamefont {Kjaergaard}\ \emph {et~al.}(2020)\citenamefont
		{Kjaergaard}, \citenamefont {Schwartz}, \citenamefont {Braum{\"u}ller},
		\citenamefont {Krantz}, \citenamefont {Wang}, \citenamefont {Gustavsson},\
		and\ \citenamefont {Oliver}}]{kjaergaard2020superconducting}%
	\BibitemOpen
	\bibfield  {author} {\bibinfo {author} {\bibfnamefont {M.}~\bibnamefont
			{Kjaergaard}}, \bibinfo {author} {\bibfnamefont {M.~E.}\ \bibnamefont
			{Schwartz}}, \bibinfo {author} {\bibfnamefont {J.}~\bibnamefont
			{Braum{\"u}ller}}, \bibinfo {author} {\bibfnamefont {P.}~\bibnamefont
			{Krantz}}, \bibinfo {author} {\bibfnamefont {J.~I.-J.}\ \bibnamefont {Wang}},
		\bibinfo {author} {\bibfnamefont {S.}~\bibnamefont {Gustavsson}},\ and\
		\bibinfo {author} {\bibfnamefont {W.~D.}\ \bibnamefont {Oliver}},\
	}\href@noop {} {\bibfield  {journal} {\bibinfo  {journal} {Annual Review of
				Condensed Matter Physics}\ }\textbf {\bibinfo {volume} {11}},\ \bibinfo
		{pages} {369} (\bibinfo {year} {2020})}\BibitemShut {NoStop}%
	\bibitem [{\citenamefont {Madsen}\ \emph {et~al.}(2004)\citenamefont {Madsen},
		\citenamefont {Hensinger}, \citenamefont {Stick}, \citenamefont {Rabchuk},\
		and\ \citenamefont {Monroe}}]{madsen2004planar}%
	\BibitemOpen
	\bibfield  {author} {\bibinfo {author} {\bibfnamefont {M.}~\bibnamefont
			{Madsen}}, \bibinfo {author} {\bibfnamefont {W.}~\bibnamefont {Hensinger}},
		\bibinfo {author} {\bibfnamefont {D.}~\bibnamefont {Stick}}, \bibinfo
		{author} {\bibfnamefont {J.}~\bibnamefont {Rabchuk}},\ and\ \bibinfo {author}
		{\bibfnamefont {C.}~\bibnamefont {Monroe}},\ }\href@noop {} {\bibfield
		{journal} {\bibinfo  {journal} {Applied Physics B}\ }\textbf {\bibinfo
			{volume} {78}},\ \bibinfo {pages} {639} (\bibinfo {year} {2004})}\BibitemShut
	{NoStop}%
	\bibitem [{\citenamefont {Pearson}\ \emph {et~al.}(2006)\citenamefont
		{Pearson}, \citenamefont {Leibrandt}, \citenamefont {Bakr}, \citenamefont
		{Mallard}, \citenamefont {Brown},\ and\ \citenamefont
		{Chuang}}]{pearson2006experimental}%
	\BibitemOpen
	\bibfield  {author} {\bibinfo {author} {\bibfnamefont {C.}~\bibnamefont
			{Pearson}}, \bibinfo {author} {\bibfnamefont {D.}~\bibnamefont {Leibrandt}},
		\bibinfo {author} {\bibfnamefont {W.}~\bibnamefont {Bakr}}, \bibinfo {author}
		{\bibfnamefont {W.}~\bibnamefont {Mallard}}, \bibinfo {author} {\bibfnamefont
			{K.}~\bibnamefont {Brown}},\ and\ \bibinfo {author} {\bibfnamefont
			{I.}~\bibnamefont {Chuang}},\ }\href@noop {} {\bibfield  {journal} {\bibinfo
			{journal} {Physical Review A}\ }\textbf {\bibinfo {volume} {73}},\ \bibinfo
		{pages} {032307} (\bibinfo {year} {2006})}\BibitemShut {NoStop}%
	\bibitem [{\citenamefont {Kim}\ \emph {et~al.}(2010)\citenamefont {Kim},
		\citenamefont {Herskind}, \citenamefont {Kim}, \citenamefont {Kim},\ and\
		\citenamefont {Chuang}}]{kim2010surface}%
	\BibitemOpen
	\bibfield  {author} {\bibinfo {author} {\bibfnamefont {T.~H.}\ \bibnamefont
			{Kim}}, \bibinfo {author} {\bibfnamefont {P.~F.}\ \bibnamefont {Herskind}},
		\bibinfo {author} {\bibfnamefont {T.}~\bibnamefont {Kim}}, \bibinfo {author}
		{\bibfnamefont {J.}~\bibnamefont {Kim}},\ and\ \bibinfo {author}
		{\bibfnamefont {I.~L.}\ \bibnamefont {Chuang}},\ }\href@noop {} {\bibfield
		{journal} {\bibinfo  {journal} {Physical Review A}\ }\textbf {\bibinfo
			{volume} {82}},\ \bibinfo {pages} {043412} (\bibinfo {year}
		{2010})}\BibitemShut {NoStop}%
	\bibitem [{\citenamefont {Latorre}\ \emph {et~al.}(2022)\citenamefont
		{Latorre}, \citenamefont {Paradkar}, \citenamefont {Hambraeus}, \citenamefont
		{Higgins},\ and\ \citenamefont {Wieczorek}}]{latorre2022chip}%
	\BibitemOpen
	\bibfield  {author} {\bibinfo {author} {\bibfnamefont {M.~G.}\ \bibnamefont
			{Latorre}}, \bibinfo {author} {\bibfnamefont {A.}~\bibnamefont {Paradkar}},
		\bibinfo {author} {\bibfnamefont {D.}~\bibnamefont {Hambraeus}}, \bibinfo
		{author} {\bibfnamefont {G.}~\bibnamefont {Higgins}},\ and\ \bibinfo {author}
		{\bibfnamefont {W.}~\bibnamefont {Wieczorek}},\ }\href@noop {} {\bibfield
		{journal} {\bibinfo  {journal} {IEEE Transactions on Applied
				Superconductivity}\ }\textbf {\bibinfo {volume} {32}},\ \bibinfo {pages} {1}
		(\bibinfo {year} {2022})}\BibitemShut {NoStop}%
	\bibitem [{\citenamefont {Magrini}\ \emph {et~al.}(2018)\citenamefont
		{Magrini}, \citenamefont {Norte}, \citenamefont {Riedinger}, \citenamefont
		{Marinkovi{\'c}}, \citenamefont {Grass}, \citenamefont {Deli{\'c}},
		\citenamefont {Gr{\"o}blacher}, \citenamefont {Hong},\ and\ \citenamefont
		{Aspelmeyer}}]{magrini2018near}%
	\BibitemOpen
	\bibfield  {author} {\bibinfo {author} {\bibfnamefont {L.}~\bibnamefont
			{Magrini}}, \bibinfo {author} {\bibfnamefont {R.~A.}\ \bibnamefont {Norte}},
		\bibinfo {author} {\bibfnamefont {R.}~\bibnamefont {Riedinger}}, \bibinfo
		{author} {\bibfnamefont {I.}~\bibnamefont {Marinkovi{\'c}}}, \bibinfo
		{author} {\bibfnamefont {D.}~\bibnamefont {Grass}}, \bibinfo {author}
		{\bibfnamefont {U.}~\bibnamefont {Deli{\'c}}}, \bibinfo {author}
		{\bibfnamefont {S.}~\bibnamefont {Gr{\"o}blacher}}, \bibinfo {author}
		{\bibfnamefont {S.}~\bibnamefont {Hong}},\ and\ \bibinfo {author}
		{\bibfnamefont {M.}~\bibnamefont {Aspelmeyer}},\ }\href@noop {} {\bibfield
		{journal} {\bibinfo  {journal} {Optica}\ }\textbf {\bibinfo {volume} {5}},\
		\bibinfo {pages} {1597} (\bibinfo {year} {2018})}\BibitemShut {NoStop}%
	\bibitem [{\citenamefont {Rieser}\ \emph {et~al.}(2022)\citenamefont {Rieser},
		\citenamefont {Ciampini}, \citenamefont {Rudolph}, \citenamefont {Kiesel},
		\citenamefont {Hornberger}, \citenamefont {Stickler}, \citenamefont
		{Aspelmeyer},\ and\ \citenamefont {Deli{\'c}}}]{rieser2022tunable}%
	\BibitemOpen
	\bibfield  {author} {\bibinfo {author} {\bibfnamefont {J.}~\bibnamefont
			{Rieser}}, \bibinfo {author} {\bibfnamefont {M.~A.}\ \bibnamefont
			{Ciampini}}, \bibinfo {author} {\bibfnamefont {H.}~\bibnamefont {Rudolph}},
		\bibinfo {author} {\bibfnamefont {N.}~\bibnamefont {Kiesel}}, \bibinfo
		{author} {\bibfnamefont {K.}~\bibnamefont {Hornberger}}, \bibinfo {author}
		{\bibfnamefont {B.~A.}\ \bibnamefont {Stickler}}, \bibinfo {author}
		{\bibfnamefont {M.}~\bibnamefont {Aspelmeyer}},\ and\ \bibinfo {author}
		{\bibfnamefont {U.}~\bibnamefont {Deli{\'c}}},\ }\href@noop {} {\bibfield
		{journal} {\bibinfo  {journal} {Science}\ }\textbf {\bibinfo {volume}
			{377}},\ \bibinfo {pages} {987} (\bibinfo {year} {2022})}\BibitemShut
	{NoStop}%
	\bibitem [{\citenamefont {Vijayan}\ \emph {et~al.}(2023)\citenamefont
		{Vijayan}, \citenamefont {Zhang}, \citenamefont {Piotrowski}, \citenamefont
		{Windey}, \citenamefont {van~der Laan}, \citenamefont {Frimmer},\ and\
		\citenamefont {Novotny}}]{vijayan2023scalable}%
	\BibitemOpen
	\bibfield  {author} {\bibinfo {author} {\bibfnamefont {J.}~\bibnamefont
			{Vijayan}}, \bibinfo {author} {\bibfnamefont {Z.}~\bibnamefont {Zhang}},
		\bibinfo {author} {\bibfnamefont {J.}~\bibnamefont {Piotrowski}}, \bibinfo
		{author} {\bibfnamefont {D.}~\bibnamefont {Windey}}, \bibinfo {author}
		{\bibfnamefont {F.}~\bibnamefont {van~der Laan}}, \bibinfo {author}
		{\bibfnamefont {M.}~\bibnamefont {Frimmer}},\ and\ \bibinfo {author}
		{\bibfnamefont {L.}~\bibnamefont {Novotny}},\ }\href@noop {} {\bibfield
		{journal} {\bibinfo  {journal} {Nature Nanotechnology}\ }\textbf {\bibinfo
			{volume} {18}},\ \bibinfo {pages} {49} (\bibinfo {year} {2023})}\BibitemShut
	{NoStop}%
	\bibitem [{\citenamefont {{\v{C}}i{\v{z}}m{\'a}r}\ \emph
		{et~al.}(2006)\citenamefont {{\v{C}}i{\v{z}}m{\'a}r}, \citenamefont
		{{\v{S}}iler}, \citenamefont {{\v{S}}er{\`y}}, \citenamefont {Zem{\'a}nek},
		\citenamefont {Garc{\'e}s-Ch{\'a}vez},\ and\ \citenamefont
		{Dholakia}}]{vcivzmar2006optical}%
	\BibitemOpen
	\bibfield  {author} {\bibinfo {author} {\bibfnamefont {T.}~\bibnamefont
			{{\v{C}}i{\v{z}}m{\'a}r}}, \bibinfo {author} {\bibfnamefont {M.}~\bibnamefont
			{{\v{S}}iler}}, \bibinfo {author} {\bibfnamefont {M.}~\bibnamefont
			{{\v{S}}er{\`y}}}, \bibinfo {author} {\bibfnamefont {P.}~\bibnamefont
			{Zem{\'a}nek}}, \bibinfo {author} {\bibfnamefont {V.}~\bibnamefont
			{Garc{\'e}s-Ch{\'a}vez}},\ and\ \bibinfo {author} {\bibfnamefont
			{K.}~\bibnamefont {Dholakia}},\ }\href@noop {} {\bibfield  {journal}
		{\bibinfo  {journal} {Physical Review B}\ }\textbf {\bibinfo {volume} {74}},\
		\bibinfo {pages} {035105} (\bibinfo {year} {2006})}\BibitemShut {NoStop}%
	\bibitem [{\citenamefont {Roda-Llordes}\ \emph {et~al.}(2023)\citenamefont
		{Roda-Llordes}, \citenamefont {Riera-Campeny}, \citenamefont {Candoli},
		\citenamefont {Grochowski},\ and\ \citenamefont
		{Romero-Isart}}]{roda2023macroscopic}%
	\BibitemOpen
	\bibfield  {author} {\bibinfo {author} {\bibfnamefont {M.}~\bibnamefont
			{Roda-Llordes}}, \bibinfo {author} {\bibfnamefont {A.}~\bibnamefont
			{Riera-Campeny}}, \bibinfo {author} {\bibfnamefont {D.}~\bibnamefont
			{Candoli}}, \bibinfo {author} {\bibfnamefont {P.~T.}\ \bibnamefont
			{Grochowski}},\ and\ \bibinfo {author} {\bibfnamefont {O.}~\bibnamefont
			{Romero-Isart}},\ }\href@noop {} {\bibfield  {journal} {\bibinfo  {journal}
			{arXiv preprint arXiv:2303.07959}\ } (\bibinfo {year} {2023})}\BibitemShut
	{NoStop}%
	\bibitem [{\citenamefont {Jin}\ \emph {et~al.}(2023)\citenamefont {Jin},
		\citenamefont {Shen}, \citenamefont {Ju}, \citenamefont {Gao}, \citenamefont
		{Zu}, \citenamefont {Grine},\ and\ \citenamefont {Li}}]{jin2023quantum}%
	\BibitemOpen
	\bibfield  {author} {\bibinfo {author} {\bibfnamefont {Y.}~\bibnamefont
			{Jin}}, \bibinfo {author} {\bibfnamefont {K.}~\bibnamefont {Shen}}, \bibinfo
		{author} {\bibfnamefont {P.}~\bibnamefont {Ju}}, \bibinfo {author}
		{\bibfnamefont {X.}~\bibnamefont {Gao}}, \bibinfo {author} {\bibfnamefont
			{C.}~\bibnamefont {Zu}}, \bibinfo {author} {\bibfnamefont {A.~J.}\
			\bibnamefont {Grine}},\ and\ \bibinfo {author} {\bibfnamefont
			{T.}~\bibnamefont {Li}},\ }\href@noop {} {\bibfield  {journal} {\bibinfo
			{journal} {arXiv preprint arXiv:2309.05821}\ } (\bibinfo {year}
		{2023})}\BibitemShut {NoStop}%
	\bibitem [{\citenamefont {Alda}\ \emph {et~al.}(2016)\citenamefont {Alda},
		\citenamefont {Berthelot}, \citenamefont {Rica},\ and\ \citenamefont
		{Quidant}}]{alda2016trapping}%
	\BibitemOpen
	\bibfield  {author} {\bibinfo {author} {\bibfnamefont {I.}~\bibnamefont
			{Alda}}, \bibinfo {author} {\bibfnamefont {J.}~\bibnamefont {Berthelot}},
		\bibinfo {author} {\bibfnamefont {R.~A.}\ \bibnamefont {Rica}},\ and\
		\bibinfo {author} {\bibfnamefont {R.}~\bibnamefont {Quidant}},\ }\href@noop
	{} {\bibfield  {journal} {\bibinfo  {journal} {Applied Physics Letters}\
		}\textbf {\bibinfo {volume} {109}} (\bibinfo {year} {2016})}\BibitemShut
	{NoStop}%
	\bibitem [{\citenamefont {Shen}\ \emph {et~al.}(2021)\citenamefont {Shen},
		\citenamefont {Duan}, \citenamefont {Ju}, \citenamefont {Xu}, \citenamefont
		{Chen}, \citenamefont {Zhang}, \citenamefont {Ahn}, \citenamefont {Ni},\ and\
		\citenamefont {Li}}]{shen2021chip}%
	\BibitemOpen
	\bibfield  {author} {\bibinfo {author} {\bibfnamefont {K.}~\bibnamefont
			{Shen}}, \bibinfo {author} {\bibfnamefont {Y.}~\bibnamefont {Duan}}, \bibinfo
		{author} {\bibfnamefont {P.}~\bibnamefont {Ju}}, \bibinfo {author}
		{\bibfnamefont {Z.}~\bibnamefont {Xu}}, \bibinfo {author} {\bibfnamefont
			{X.}~\bibnamefont {Chen}}, \bibinfo {author} {\bibfnamefont {L.}~\bibnamefont
			{Zhang}}, \bibinfo {author} {\bibfnamefont {J.}~\bibnamefont {Ahn}}, \bibinfo
		{author} {\bibfnamefont {X.}~\bibnamefont {Ni}},\ and\ \bibinfo {author}
		{\bibfnamefont {T.}~\bibnamefont {Li}},\ }\href@noop {} {\bibfield  {journal}
		{\bibinfo  {journal} {Optica}\ }\textbf {\bibinfo {volume} {8}},\ \bibinfo
		{pages} {1359} (\bibinfo {year} {2021})}\BibitemShut {NoStop}%
	\bibitem [{\citenamefont {Zem{\'a}nek}\ \emph {et~al.}(1998)\citenamefont
		{Zem{\'a}nek}, \citenamefont {Jon{\'a}{\v{s}}}, \citenamefont
		{{\v{S}}r{\'a}mek},\ and\ \citenamefont {Li{\v{s}}ka}}]{zemanek1998optical}%
	\BibitemOpen
	\bibfield  {author} {\bibinfo {author} {\bibfnamefont {P.}~\bibnamefont
			{Zem{\'a}nek}}, \bibinfo {author} {\bibfnamefont {A.}~\bibnamefont
			{Jon{\'a}{\v{s}}}}, \bibinfo {author} {\bibfnamefont {L.}~\bibnamefont
			{{\v{S}}r{\'a}mek}},\ and\ \bibinfo {author} {\bibfnamefont {M.}~\bibnamefont
			{Li{\v{s}}ka}},\ }\href@noop {} {\bibfield  {journal} {\bibinfo  {journal}
			{Optics communications}\ }\textbf {\bibinfo {volume} {151}},\ \bibinfo
		{pages} {273} (\bibinfo {year} {1998})}\BibitemShut {NoStop}%
	\bibitem [{\citenamefont {Kamba}\ \emph {et~al.}(2021)\citenamefont {Kamba},
		\citenamefont {Kiuchi}, \citenamefont {Yotsuya},\ and\ \citenamefont
		{Aikawa}}]{kamba2021recoil}%
	\BibitemOpen
	\bibfield  {author} {\bibinfo {author} {\bibfnamefont {M.}~\bibnamefont
			{Kamba}}, \bibinfo {author} {\bibfnamefont {H.}~\bibnamefont {Kiuchi}},
		\bibinfo {author} {\bibfnamefont {T.}~\bibnamefont {Yotsuya}},\ and\ \bibinfo
		{author} {\bibfnamefont {K.}~\bibnamefont {Aikawa}},\ }\href@noop {}
	{\bibfield  {journal} {\bibinfo  {journal} {Physical Review A}\ }\textbf
		{\bibinfo {volume} {103}},\ \bibinfo {pages} {L051701} (\bibinfo {year}
		{2021})}\BibitemShut {NoStop}%
	\bibitem [{\citenamefont {Ashkin}(1970)}]{ashkin1970acceleration}%
	\BibitemOpen
	\bibfield  {author} {\bibinfo {author} {\bibfnamefont {A.}~\bibnamefont
			{Ashkin}},\ }\href@noop {} {\bibfield  {journal} {\bibinfo  {journal}
			{Physical review letters}\ }\textbf {\bibinfo {volume} {24}},\ \bibinfo
		{pages} {156} (\bibinfo {year} {1970})}\BibitemShut {NoStop}%
	\bibitem [{\citenamefont {Constable}\ \emph {et~al.}(1993)\citenamefont
		{Constable}, \citenamefont {Kim}, \citenamefont {Mervis}, \citenamefont
		{Zarinetchi},\ and\ \citenamefont {Prentiss}}]{constable1993demonstration}%
	\BibitemOpen
	\bibfield  {author} {\bibinfo {author} {\bibfnamefont {A.}~\bibnamefont
			{Constable}}, \bibinfo {author} {\bibfnamefont {J.}~\bibnamefont {Kim}},
		\bibinfo {author} {\bibfnamefont {J.}~\bibnamefont {Mervis}}, \bibinfo
		{author} {\bibfnamefont {F.}~\bibnamefont {Zarinetchi}},\ and\ \bibinfo
		{author} {\bibfnamefont {M.}~\bibnamefont {Prentiss}},\ }\href@noop {}
	{\bibfield  {journal} {\bibinfo  {journal} {Optics letters}\ }\textbf
		{\bibinfo {volume} {18}},\ \bibinfo {pages} {1867} (\bibinfo {year}
		{1993})}\BibitemShut {NoStop}%
	\bibitem [{\citenamefont {Maurer}\ \emph {et~al.}(2023)\citenamefont {Maurer},
		\citenamefont {Gonzalez-Ballestero},\ and\ \citenamefont
		{Romero-Isart}}]{maurer2023quantum}%
	\BibitemOpen
	\bibfield  {author} {\bibinfo {author} {\bibfnamefont {P.}~\bibnamefont
			{Maurer}}, \bibinfo {author} {\bibfnamefont {C.}~\bibnamefont
			{Gonzalez-Ballestero}},\ and\ \bibinfo {author} {\bibfnamefont
			{O.}~\bibnamefont {Romero-Isart}},\ }\href@noop {} {\bibfield  {journal}
		{\bibinfo  {journal} {Physical Review A}\ }\textbf {\bibinfo {volume}
			{108}},\ \bibinfo {pages} {033714} (\bibinfo {year} {2023})}\BibitemShut
	{NoStop}%
	\bibitem [{\citenamefont {Yu}\ \emph {et~al.}()\citenamefont {Yu},
		\citenamefont {Ranno}, \citenamefont {Du}, \citenamefont {Serna},
		\citenamefont {McDonough}, \citenamefont {Fahrenkopf}, \citenamefont {Gu},\
		and\ \citenamefont {Hu}}]{yu2023free}%
	\BibitemOpen
	\bibfield  {author} {\bibinfo {author} {\bibfnamefont {S.}~\bibnamefont
			{Yu}}, \bibinfo {author} {\bibfnamefont {L.}~\bibnamefont {Ranno}}, \bibinfo
		{author} {\bibfnamefont {Q.}~\bibnamefont {Du}}, \bibinfo {author}
		{\bibfnamefont {S.}~\bibnamefont {Serna}}, \bibinfo {author} {\bibfnamefont
			{C.}~\bibnamefont {McDonough}}, \bibinfo {author} {\bibfnamefont
			{N.}~\bibnamefont {Fahrenkopf}}, \bibinfo {author} {\bibfnamefont
			{T.}~\bibnamefont {Gu}},\ and\ \bibinfo {author} {\bibfnamefont
			{J.}~\bibnamefont {Hu}},\ }\href
	{https://doi.org/https://doi.org/10.1002/lpor.202200025} {\bibfield
		{journal} {\bibinfo  {journal} {Laser \& Photonics Reviews}\ }\textbf
		{\bibinfo {volume} {17}},\ \bibinfo {pages} {2200025}}\BibitemShut {NoStop}%
	\bibitem [{\citenamefont {Wang}\ \emph {et~al.}(2023)\citenamefont {Wang},
		\citenamefont {Zhang}, \citenamefont {Ladika}, \citenamefont {Yu},
		\citenamefont {Gailevi{\v{c}}ius}, \citenamefont {Wang}, \citenamefont {Pan},
		\citenamefont {Nair}, \citenamefont {Ke}, \citenamefont {Mori} \emph
		{et~al.}}]{wang2023two}%
	\BibitemOpen
	\bibfield  {author} {\bibinfo {author} {\bibfnamefont {H.}~\bibnamefont
			{Wang}}, \bibinfo {author} {\bibfnamefont {W.}~\bibnamefont {Zhang}},
		\bibinfo {author} {\bibfnamefont {D.}~\bibnamefont {Ladika}}, \bibinfo
		{author} {\bibfnamefont {H.}~\bibnamefont {Yu}}, \bibinfo {author}
		{\bibfnamefont {D.}~\bibnamefont {Gailevi{\v{c}}ius}}, \bibinfo {author}
		{\bibfnamefont {H.}~\bibnamefont {Wang}}, \bibinfo {author} {\bibfnamefont
			{C.-F.}\ \bibnamefont {Pan}}, \bibinfo {author} {\bibfnamefont {P.~N.~S.}\
			\bibnamefont {Nair}}, \bibinfo {author} {\bibfnamefont {Y.}~\bibnamefont
			{Ke}}, \bibinfo {author} {\bibfnamefont {T.}~\bibnamefont {Mori}}, \emph
		{et~al.},\ }\href@noop {} {\bibfield  {journal} {\bibinfo  {journal}
			{Advanced Functional Materials}\ ,\ \bibinfo {pages} {2214211}} (\bibinfo
		{year} {2023})}\BibitemShut {NoStop}%
	\bibitem [{\citenamefont {Tebbenjohanns}\ \emph
		{et~al.}(2019{\natexlab{b}})\citenamefont {Tebbenjohanns}, \citenamefont
		{Frimmer},\ and\ \citenamefont {Novotny}}]{tebbenjohanns2019optimal}%
	\BibitemOpen
	\bibfield  {author} {\bibinfo {author} {\bibfnamefont {F.}~\bibnamefont
			{Tebbenjohanns}}, \bibinfo {author} {\bibfnamefont {M.}~\bibnamefont
			{Frimmer}},\ and\ \bibinfo {author} {\bibfnamefont {L.}~\bibnamefont
			{Novotny}},\ }\href@noop {} {\bibfield  {journal} {\bibinfo  {journal}
			{Physical Review A}\ }\textbf {\bibinfo {volume} {100}},\ \bibinfo {pages}
		{043821} (\bibinfo {year} {2019}{\natexlab{b}})}\BibitemShut {NoStop}%
	\bibitem [{\citenamefont {Gieseler}(2014)}]{gieseler_thesis}%
	\BibitemOpen
	\bibfield  {author} {\bibinfo {author} {\bibfnamefont {J.}~\bibnamefont
			{Gieseler}},\ }\href
	{https://www.tdx.cat/bitstream/handle/10803/144555/TJG1de1.pdf?sequence=1}
	{\emph {\bibinfo {title} {Dynamics of optically levitated nanoparticles in
				high vacuum}}}\ (\bibinfo {year} {2014})\BibitemShut {NoStop}%
	\bibitem [{\citenamefont {Hebestreit}\ \emph
		{et~al.}(2018{\natexlab{b}})\citenamefont {Hebestreit}, \citenamefont
		{Frimmer}, \citenamefont {Reimann}, \citenamefont {Dellago}, \citenamefont
		{Ricci},\ and\ \citenamefont {Novotny}}]{hebestreit_calibration_2018}%
	\BibitemOpen
	\bibfield  {author} {\bibinfo {author} {\bibfnamefont {E.}~\bibnamefont
			{Hebestreit}}, \bibinfo {author} {\bibfnamefont {M.}~\bibnamefont {Frimmer}},
		\bibinfo {author} {\bibfnamefont {R.}~\bibnamefont {Reimann}}, \bibinfo
		{author} {\bibfnamefont {C.}~\bibnamefont {Dellago}}, \bibinfo {author}
		{\bibfnamefont {F.}~\bibnamefont {Ricci}},\ and\ \bibinfo {author}
		{\bibfnamefont {L.}~\bibnamefont {Novotny}},\ }\href
	{https://doi.org/10.1063/1.5017119} {\bibfield  {journal} {\bibinfo
			{journal} {Review of Scientific Instruments}\ }\textbf {\bibinfo {volume}
			{89}},\ \bibinfo {pages} {033111} (\bibinfo {year}
		{2018}{\natexlab{b}})}\BibitemShut {NoStop}%
	\bibitem [{\citenamefont {Kubo}(1966)}]{kubo1966fluctuation}%
	\BibitemOpen
	\bibfield  {author} {\bibinfo {author} {\bibfnamefont {R.}~\bibnamefont
			{Kubo}},\ }\href@noop {} {\bibfield  {journal} {\bibinfo  {journal} {Reports
				on progress in physics}\ }\textbf {\bibinfo {volume} {29}},\ \bibinfo {pages}
		{255} (\bibinfo {year} {1966})}\BibitemShut {NoStop}%
	\bibitem [{\citenamefont {Li}(2011)}]{li2011millikelvin}%
	\BibitemOpen
	\bibfield  {author} {\bibinfo {author} {\bibfnamefont {K.~S. R.~M.}\
			\bibnamefont {Li}, \bibfnamefont {T.}},\ }\href@noop {} {\bibfield  {journal}
		{\bibinfo  {journal} {Nature Physics}\ ,\ \bibinfo {pages} {527–530}}
		(\bibinfo {year} {2011})}\BibitemShut {NoStop}%
	\bibitem [{\citenamefont {Poggio}\ \emph {et~al.}(2007)\citenamefont {Poggio},
		\citenamefont {Degen}, \citenamefont {Mamin},\ and\ \citenamefont
		{Rugar}}]{poggio2007feedback}%
	\BibitemOpen
	\bibfield  {author} {\bibinfo {author} {\bibfnamefont {M.}~\bibnamefont
			{Poggio}}, \bibinfo {author} {\bibfnamefont {C.}~\bibnamefont {Degen}},
		\bibinfo {author} {\bibfnamefont {H.}~\bibnamefont {Mamin}},\ and\ \bibinfo
		{author} {\bibfnamefont {D.}~\bibnamefont {Rugar}},\ }\href@noop {}
	{\bibfield  {journal} {\bibinfo  {journal} {Physical Review Letters}\
		}\textbf {\bibinfo {volume} {99}},\ \bibinfo {pages} {017201} (\bibinfo
		{year} {2007})}\BibitemShut {NoStop}%
	\bibitem [{\citenamefont {Burns}\ \emph {et~al.}(1989)\citenamefont {Burns},
		\citenamefont {Fournier},\ and\ \citenamefont
		{Golovchenko}}]{burns1989optical}%
	\BibitemOpen
	\bibfield  {author} {\bibinfo {author} {\bibfnamefont {M.~M.}\ \bibnamefont
			{Burns}}, \bibinfo {author} {\bibfnamefont {J.-M.}\ \bibnamefont
			{Fournier}},\ and\ \bibinfo {author} {\bibfnamefont {J.~A.}\ \bibnamefont
			{Golovchenko}},\ }\href@noop {} {\bibfield  {journal} {\bibinfo  {journal}
			{Physical Review Letters}\ }\textbf {\bibinfo {volume} {63}},\ \bibinfo
		{pages} {1233} (\bibinfo {year} {1989})}\BibitemShut {NoStop}%
	\bibitem [{\citenamefont {Lepeshov}\ \emph {et~al.}(2023)\citenamefont
		{Lepeshov}, \citenamefont {Meyer}, \citenamefont {Maurer}, \citenamefont
		{Romero-Isart},\ and\ \citenamefont {Quidant}}]{lepeshov2023levitated}%
	\BibitemOpen
	\bibfield  {author} {\bibinfo {author} {\bibfnamefont {S.}~\bibnamefont
			{Lepeshov}}, \bibinfo {author} {\bibfnamefont {N.}~\bibnamefont {Meyer}},
		\bibinfo {author} {\bibfnamefont {P.}~\bibnamefont {Maurer}}, \bibinfo
		{author} {\bibfnamefont {O.}~\bibnamefont {Romero-Isart}},\ and\ \bibinfo
		{author} {\bibfnamefont {R.}~\bibnamefont {Quidant}},\ }\href@noop {}
	{\bibfield  {journal} {\bibinfo  {journal} {Physical Review Letters}\
		}\textbf {\bibinfo {volume} {130}},\ \bibinfo {pages} {233601} (\bibinfo
		{year} {2023})}\BibitemShut {NoStop}%
	\bibitem [{\citenamefont {Windpassinger}\ and\ \citenamefont
		{Sengstock}(2013)}]{windpassinger2013engineering}%
	\BibitemOpen
	\bibfield  {author} {\bibinfo {author} {\bibfnamefont {P.}~\bibnamefont
			{Windpassinger}}\ and\ \bibinfo {author} {\bibfnamefont {K.}~\bibnamefont
			{Sengstock}},\ }\href@noop {} {\bibfield  {journal} {\bibinfo  {journal}
			{Reports on progress in physics}\ }\textbf {\bibinfo {volume} {76}},\
		\bibinfo {pages} {086401} (\bibinfo {year} {2013})}\BibitemShut {NoStop}%
	\bibitem [{\citenamefont {Li}\ \emph {et~al.}(2021)\citenamefont {Li},
		\citenamefont {Liu}, \citenamefont {Lin}, \citenamefont {Ng},\ and\
		\citenamefont {Chan}}]{li2021non}%
	\BibitemOpen
	\bibfield  {author} {\bibinfo {author} {\bibfnamefont {X.}~\bibnamefont
			{Li}}, \bibinfo {author} {\bibfnamefont {Y.}~\bibnamefont {Liu}}, \bibinfo
		{author} {\bibfnamefont {Z.}~\bibnamefont {Lin}}, \bibinfo {author}
		{\bibfnamefont {J.}~\bibnamefont {Ng}},\ and\ \bibinfo {author}
		{\bibfnamefont {C.~T.}\ \bibnamefont {Chan}},\ }\href@noop {} {\bibfield
		{journal} {\bibinfo  {journal} {Nature communications}\ }\textbf {\bibinfo
			{volume} {12}},\ \bibinfo {pages} {6597} (\bibinfo {year}
		{2021})}\BibitemShut {NoStop}%
	\bibitem [{\citenamefont {Grass}\ \emph {et~al.}(2016)\citenamefont {Grass},
		\citenamefont {Fesel}, \citenamefont {Hofer}, \citenamefont {Kiesel},\ and\
		\citenamefont {Aspelmeyer}}]{grass2016optical}%
	\BibitemOpen
	\bibfield  {author} {\bibinfo {author} {\bibfnamefont {D.}~\bibnamefont
			{Grass}}, \bibinfo {author} {\bibfnamefont {J.}~\bibnamefont {Fesel}},
		\bibinfo {author} {\bibfnamefont {S.~G.}\ \bibnamefont {Hofer}}, \bibinfo
		{author} {\bibfnamefont {N.}~\bibnamefont {Kiesel}},\ and\ \bibinfo {author}
		{\bibfnamefont {M.}~\bibnamefont {Aspelmeyer}},\ }\href@noop {} {\bibfield
		{journal} {\bibinfo  {journal} {Applied Physics Letters}\ }\textbf {\bibinfo
			{volume} {108}} (\bibinfo {year} {2016})}\BibitemShut {NoStop}%
	\bibitem [{\citenamefont {Bykov}\ \emph {et~al.}(2018)\citenamefont {Bykov},
		\citenamefont {Xie}, \citenamefont {Zeltner}, \citenamefont {Machnev},
		\citenamefont {Wong}, \citenamefont {Euser},\ and\ \citenamefont
		{Russell}}]{bykov2018long}%
	\BibitemOpen
	\bibfield  {author} {\bibinfo {author} {\bibfnamefont {D.~S.}\ \bibnamefont
			{Bykov}}, \bibinfo {author} {\bibfnamefont {S.}~\bibnamefont {Xie}}, \bibinfo
		{author} {\bibfnamefont {R.}~\bibnamefont {Zeltner}}, \bibinfo {author}
		{\bibfnamefont {A.}~\bibnamefont {Machnev}}, \bibinfo {author} {\bibfnamefont
			{G.~K.}\ \bibnamefont {Wong}}, \bibinfo {author} {\bibfnamefont {T.~G.}\
			\bibnamefont {Euser}},\ and\ \bibinfo {author} {\bibfnamefont {P.~S.~J.}\
			\bibnamefont {Russell}},\ }\href@noop {} {\bibfield  {journal} {\bibinfo
			{journal} {Light: Science \& Applications}\ }\textbf {\bibinfo {volume}
			{7}},\ \bibinfo {pages} {22} (\bibinfo {year} {2018})}\BibitemShut {NoStop}%
	\bibitem [{\citenamefont {Stickler}\ \emph {et~al.}(2018)\citenamefont
		{Stickler}, \citenamefont {Papendell}, \citenamefont {Kuhn}, \citenamefont
		{Schrinski}, \citenamefont {Millen}, \citenamefont {Arndt},\ and\
		\citenamefont {Hornberger}}]{stickler2018probing}%
	\BibitemOpen
	\bibfield  {author} {\bibinfo {author} {\bibfnamefont {B.~A.}\ \bibnamefont
			{Stickler}}, \bibinfo {author} {\bibfnamefont {B.}~\bibnamefont {Papendell}},
		\bibinfo {author} {\bibfnamefont {S.}~\bibnamefont {Kuhn}}, \bibinfo {author}
		{\bibfnamefont {B.}~\bibnamefont {Schrinski}}, \bibinfo {author}
		{\bibfnamefont {J.}~\bibnamefont {Millen}}, \bibinfo {author} {\bibfnamefont
			{M.}~\bibnamefont {Arndt}},\ and\ \bibinfo {author} {\bibfnamefont
			{K.}~\bibnamefont {Hornberger}},\ }\href@noop {} {\bibfield  {journal}
		{\bibinfo  {journal} {New Journal of Physics}\ }\textbf {\bibinfo {volume}
			{20}},\ \bibinfo {pages} {122001} (\bibinfo {year} {2018})}\BibitemShut
	{NoStop}%
	\bibitem [{\citenamefont {Gonzalez-Hernandez}\ \emph
		{et~al.}(2023)\citenamefont {Gonzalez-Hernandez}, \citenamefont
		{Varapnickas}, \citenamefont {Bertoncini}, \citenamefont {Liberale},\ and\
		\citenamefont {Malinauskas}}]{gonzalez2023micro}%
	\BibitemOpen
	\bibfield  {author} {\bibinfo {author} {\bibfnamefont {D.}~\bibnamefont
			{Gonzalez-Hernandez}}, \bibinfo {author} {\bibfnamefont {S.}~\bibnamefont
			{Varapnickas}}, \bibinfo {author} {\bibfnamefont {A.}~\bibnamefont
			{Bertoncini}}, \bibinfo {author} {\bibfnamefont {C.}~\bibnamefont
			{Liberale}},\ and\ \bibinfo {author} {\bibfnamefont {M.}~\bibnamefont
			{Malinauskas}},\ }\href@noop {} {\bibfield  {journal} {\bibinfo  {journal}
			{Advanced Optical Materials}\ }\textbf {\bibinfo {volume} {11}},\ \bibinfo
		{pages} {2201701} (\bibinfo {year} {2023})}\BibitemShut {NoStop}%
	\bibitem [{\citenamefont {Plidschun}\ \emph {et~al.}(2021)\citenamefont
		{Plidschun}, \citenamefont {Ren}, \citenamefont {Kim}, \citenamefont
		{F{\"o}rster}, \citenamefont {Maier},\ and\ \citenamefont
		{Schmidt}}]{plidschun2021ultrahigh}%
	\BibitemOpen
	\bibfield  {author} {\bibinfo {author} {\bibfnamefont {M.}~\bibnamefont
			{Plidschun}}, \bibinfo {author} {\bibfnamefont {H.}~\bibnamefont {Ren}},
		\bibinfo {author} {\bibfnamefont {J.}~\bibnamefont {Kim}}, \bibinfo {author}
		{\bibfnamefont {R.}~\bibnamefont {F{\"o}rster}}, \bibinfo {author}
		{\bibfnamefont {S.~A.}\ \bibnamefont {Maier}},\ and\ \bibinfo {author}
		{\bibfnamefont {M.~A.}\ \bibnamefont {Schmidt}},\ }\href@noop {} {\bibfield
		{journal} {\bibinfo  {journal} {Light: Science \& Applications}\ }\textbf
		{\bibinfo {volume} {10}},\ \bibinfo {pages} {57} (\bibinfo {year}
		{2021})}\BibitemShut {NoStop}%
	\bibitem [{\citenamefont {Ren}\ \emph {et~al.}(2022)\citenamefont {Ren},
		\citenamefont {Jang}, \citenamefont {Li}, \citenamefont {Aigner},
		\citenamefont {Plidschun}, \citenamefont {Kim}, \citenamefont {Rho},
		\citenamefont {Schmidt},\ and\ \citenamefont {Maier}}]{ren2022achromatic}%
	\BibitemOpen
	\bibfield  {author} {\bibinfo {author} {\bibfnamefont {H.}~\bibnamefont
			{Ren}}, \bibinfo {author} {\bibfnamefont {J.}~\bibnamefont {Jang}}, \bibinfo
		{author} {\bibfnamefont {C.}~\bibnamefont {Li}}, \bibinfo {author}
		{\bibfnamefont {A.}~\bibnamefont {Aigner}}, \bibinfo {author} {\bibfnamefont
			{M.}~\bibnamefont {Plidschun}}, \bibinfo {author} {\bibfnamefont
			{J.}~\bibnamefont {Kim}}, \bibinfo {author} {\bibfnamefont {J.}~\bibnamefont
			{Rho}}, \bibinfo {author} {\bibfnamefont {M.~A.}\ \bibnamefont {Schmidt}},\
		and\ \bibinfo {author} {\bibfnamefont {S.~A.}\ \bibnamefont {Maier}},\
	}\href@noop {} {\bibfield  {journal} {\bibinfo  {journal} {nature
				communications}\ }\textbf {\bibinfo {volume} {13}},\ \bibinfo {pages} {4183}
		(\bibinfo {year} {2022})}\BibitemShut {NoStop}%
	\bibitem [{\citenamefont {Steinmetz}\ \emph {et~al.}(2006)\citenamefont
		{Steinmetz}, \citenamefont {Colombe}, \citenamefont {Hunger}, \citenamefont
		{H{\"a}nsch}, \citenamefont {Balocchi}, \citenamefont {Warburton},\ and\
		\citenamefont {Reichel}}]{steinmetz2006stable}%
	\BibitemOpen
	\bibfield  {author} {\bibinfo {author} {\bibfnamefont {T.}~\bibnamefont
			{Steinmetz}}, \bibinfo {author} {\bibfnamefont {Y.}~\bibnamefont {Colombe}},
		\bibinfo {author} {\bibfnamefont {D.}~\bibnamefont {Hunger}}, \bibinfo
		{author} {\bibfnamefont {T.}~\bibnamefont {H{\"a}nsch}}, \bibinfo {author}
		{\bibfnamefont {A.}~\bibnamefont {Balocchi}}, \bibinfo {author}
		{\bibfnamefont {R.}~\bibnamefont {Warburton}},\ and\ \bibinfo {author}
		{\bibfnamefont {J.}~\bibnamefont {Reichel}},\ }\href@noop {} {\bibfield
		{journal} {\bibinfo  {journal} {Applied Physics Letters}\ }\textbf {\bibinfo
			{volume} {89}} (\bibinfo {year} {2006})}\BibitemShut {NoStop}%
	\bibitem [{\citenamefont {Hunger}\ \emph {et~al.}(2010)\citenamefont {Hunger},
		\citenamefont {Steinmetz}, \citenamefont {Colombe}, \citenamefont {Deutsch},
		\citenamefont {H{\"a}nsch},\ and\ \citenamefont {Reichel}}]{hunger2010fiber}%
	\BibitemOpen
	\bibfield  {author} {\bibinfo {author} {\bibfnamefont {D.}~\bibnamefont
			{Hunger}}, \bibinfo {author} {\bibfnamefont {T.}~\bibnamefont {Steinmetz}},
		\bibinfo {author} {\bibfnamefont {Y.}~\bibnamefont {Colombe}}, \bibinfo
		{author} {\bibfnamefont {C.}~\bibnamefont {Deutsch}}, \bibinfo {author}
		{\bibfnamefont {T.~W.}\ \bibnamefont {H{\"a}nsch}},\ and\ \bibinfo {author}
		{\bibfnamefont {J.}~\bibnamefont {Reichel}},\ }\href@noop {} {\bibfield
		{journal} {\bibinfo  {journal} {New Journal of Physics}\ }\textbf {\bibinfo
			{volume} {12}},\ \bibinfo {pages} {065038} (\bibinfo {year}
		{2010})}\BibitemShut {NoStop}%
	\bibitem [{\citenamefont {Muller}\ \emph {et~al.}(2010)\citenamefont {Muller},
		\citenamefont {Flagg}, \citenamefont {Lawall},\ and\ \citenamefont
		{Solomon}}]{muller2010ultrahigh}%
	\BibitemOpen
	\bibfield  {author} {\bibinfo {author} {\bibfnamefont {A.}~\bibnamefont
			{Muller}}, \bibinfo {author} {\bibfnamefont {E.~B.}\ \bibnamefont {Flagg}},
		\bibinfo {author} {\bibfnamefont {J.~R.}\ \bibnamefont {Lawall}},\ and\
		\bibinfo {author} {\bibfnamefont {G.~S.}\ \bibnamefont {Solomon}},\
	}\href@noop {} {\bibfield  {journal} {\bibinfo  {journal} {Optics letters}\
		}\textbf {\bibinfo {volume} {35}},\ \bibinfo {pages} {2293} (\bibinfo {year}
		{2010})}\BibitemShut {NoStop}%
\end{thebibliography}

%

\end{document}